\documentclass[aps,prd,nofootinbib,floatfix,showpacs,notitlepage,eqsecnum,twocolumn,10pt]{revtex4-2}

\usepackage{graphicx,wrapfig,slashed,bbold,bm}
\usepackage{amsmath,amssymb,epsfig,graphicx}
\usepackage[dvipsnames]{xcolor}
\usepackage{braket}
\usepackage{epstopdf}
\usepackage{multirow}
\usepackage{booktabs}
\usepackage{ragged2e}
\usepackage{mciteplus}
\usepackage[colorlinks=true,linkcolor=blue,urlcolor=blue,citecolor=blue]{hyperref}
\usepackage{cleveref}
\usepackage{simpler-wick}
\usepackage{mathrsfs}
\usepackage{slashed}

\usepackage{chngcntr}
\counterwithout{equation}{section}

\usepackage{cancel}
\usepackage[normalem]{ulem}
\usepackage{makecell}
\usepackage{tikz}
\newcommand*\circled[1]{\tikz[baseline=(char.base)]{
            \node[shape=circle,draw,inner sep=1pt] (char) {#1};}}

\newcommand{\nc}{\newcommand}
\nc{\non}{\nonumber}
\nc{\hc}{\hbox {H.c.}}
\nc{\noi}{\noindent}
\nc{\barx}{\bar{x}}
\nc{\pbarn}{\;\hbox {pb}}
\nc{\fbarn}{\;\hbox {fb}}
\nc{\cone}{{\scriptsize \textcolor{orange}{\circled{1}}}}
\nc{\ctwo}{{\scriptsize \textcolor{orange}{\circled{2}}}}
\nc{\cthr}{{\scriptsize \textcolor{orange}{\circled{3}}}}

\nc{\hsp}{\hspace{0.5cm}}
\nc{\lsp}{\hspace{1cm}}
\nc{\Lsp}{\hspace{2cm}}
\nc{\LLsp}{\lsp\lsp}
\nc{\lra}{\longrightarrow}
\nc{\p}{\prime}
\nc{\sgn}{\text{sgn}}
\nc{\ph}{\varphi}
\nc{\op}{{\cal O}}
\nc{\tr}{\mathrm{tr}}
\nc{\eff}{\mathrm{eff}}
\nc{\sqM}{\sqrt{M}}
\nc{\NL}{\mathrm{NL}}
\nc{\moc}{\mathcal{M}} 
\nc{\rd}[1]{\textcolor{red}{#1}}
\nc{\bl}[1]{\textcolor{blue}{#1}}

\nc{\pbar}{\bar{\psi}}
\nc{\beq}{\begin{equation}}  \nc{\eeq}{\end{equation}}
\nc{\bea}{\begin{eqnarray}}  \nc{\eea}{\end{eqnarray}}
\nc{\baa}{\begin{array}}     \nc{\eaa}{\end{array}}
\nc{\bit}{\begin{itemize}}   \nc{\eit}{\end{itemize}}
\nc{\ben}{\begin{enumerate}} \nc{\een}{\end{enumerate}}
\nc{\bce}{\begin{center}}    \nc{\ece}{\end{center}}
\nc{\bpm}{\begin{pmatrix}}   \nc{\epm}{\end{pmatrix}}
\nc{\bvt}{\begin{verbatim}}  \nc{\evt}{\end{verbatim}}
\nc{\vp}[1]{\mathbf{p}_{#1}}
\nc{\vk}{{\bm k}_\perp}
\nc{\vkp}{{\bm k}'_\perp}
\nc{\vq}{{\bm q}_\perp}
\nc{\uu}{{\uparrow\uparrow}}
\nc{\ud}{{\uparrow\downarrow}}
\nc{\du}{{\downarrow\uparrow}}
\nc{\dd}{{\downarrow\downarrow}}

\def\lsim{\mathrel{\raise.3ex\hbox{$<$\kern-.75em\lower1ex\hbox{$\sim$}}}}
\def\gsim{\mathrel{\raise.3ex\hbox{$>$\kern-.75em\lower1ex\hbox{$\sim$}}}}

\def\udots{\mathinner{\mkern1mu\raise1pt\vbox{\kern7pt\hbox{.}}\mkern2mu\raise4pt\hbox{.}\mkern2mu\raise7pt\hbox{.}\mkern1mu}}

\def\ep{\epsilon}

\def\vp{{\bf p}}

\def\al{\alpha}

\def\be{\begin{equation}}
\def\ee{\end{equation}}
\def\bea{\begin{eqnarray}}
\def\eea{\end{eqnarray}}

\def\ep{\epsilon}

\def\vp{{\bf p}}

\def\al{\alpha}

\def\be{\begin{equation}}
\def\ee{\end{equation}}
\def\bea{\begin{eqnarray}}
\def\eea{\end{eqnarray}}

\allowdisplaybreaks

\begin{document}
\preprint{\begin{flushright}
Preprint number here\\  
\end{flushright}}


\author{Andrew Lundeen}
\email[ E-mail: ]{ajlundee@ncsu.edu}
\author{Chueng-Ryong Ji}%
\email[ E-mail: ]{crji@ncsu.edu}
\affiliation{%
 Department of Physics, North Carolina State University,
 Raleigh, North Carolina 27695-8202, USA
}%

\author{Yongwoo Choi}
\email[ E-mail: ]{sunctchoi@gmail.com}
\affiliation{Physics Research Institute, Inha University, Incheon 22212, Republic of Korea}
\affiliation{The Center for High Energy Physics, Kyungpook National University, Daegu 41566, Korea}

\author{Ho-Meoyng Choi}
\email[E-mail: ]{ homyoung@knu.ac.kr}
\affiliation{Department of Physics Education, Teachers College, Kyungpook National University, Daegu 41566, Korea}

\date{\today}

\title{The (3+1)-dimensional scalar field model analysis of beam spin asymmetry in the electroproduction of a scalar meson off a scalar target}

\begin{abstract}
We explore exclusive scalar meson electroproduction off a scalar target in the (3+1)-dimensional scalar field model.  This model analysis is a straightforward extension of the previous (1+1)-dimensional model analysis presented in Phys. Rev. D \textbf{105}, 096014 (2022). In contrast to the (1+1)-dimensional model, the (3+1)-dimensional model allows us to compute the beam spin asymmetry (BSA), which is proportional to the imaginary part of the product of the two Compton form factors (CFFs) that appear in the hadronic current of the present scalar meson electroproduction process. We compute both real and imaginary parts of the CFFs and note that the BSA is detectable for $-t/Q^2 \gtrsim 0.1$ although it gets quite small in the kinematic region $-t/Q^2 \ll 0.1$ where the factorization of the generalized parton distribution (GPD) is attainable. We find the analytic forms of the leading twist GPD for the DGLAP and ERBL regions in the (3+1)-dimensional scalar field model, confirming its uniqueness independent of the hadronic current component. While we verify that the GPD sum rule for the total result of summing the DGLAP and ERBL regions holds for all components of the hadronic current, we note that the respective correspondence of the DGLAP and ERBL regions to the valence and non-valence parts of the electromagnetic form factor holds only for the light-front plus component of the hadronic current but not for any other components of the hadronic current.   
We discuss the polynomiality of the GPD up to the second moments and remark on accessible ranges of kinematics to measure the BSA and CFFs with respect to the future experimental efforts of extracting the leading-twist GPDs.
\end{abstract}

\maketitle
\flushbottom
\section{INTRODUCTION}
Current and planned facilities such as Jefferson Laboratory's 12 GeV upgrade~\cite{Burkert2020,Kum2016,CLAS17b} and the future Electron-Ion Collider (EIC)~\cite{Abdul2022} are designed to map the multidimensional quark and gluon structure of hadrons. 
Exclusive meson electroproduction plays a central role in this program, providing access to Compton form factors (CFFs)~\cite{Bakker2019}, 
beam spin asymmetries (BSAs), generalized parton distributions (GPDs)~\cite{Ji:1996nm,JiMel97,Ji_1998,Radyushkin:1996nd,Rad1997,RAD99,Muel1994,PW99,Diehl:2003ny,Belitsky_2005,CJL01,CJL02}, 
electromagnetic form factors (EMFs), and gravitational form factors (GFFs)~\cite{BR08,PS18,HS17,CSC25}
that encode the fundamental information of hadron dynamics in QCD.

In this lepton-hadron scattering process, the exchanged virtual photon produces a meson off the target. 
A salient advantage of this process is the absence of the Bethe-Heitler (BH) mechanism that complicates 
deeply virtual Compton scattering (DVCS). Without involving the BH amplitude, the reaction factorizes~\cite{Collins:1996fb} rather cleanly into leptonic and hadronic currents, which enables the underlying hadronic correlation amplitudes to be accessed with minimal contamination.

The coherent electroproduction of pseudoscalar ($J^{PC}=0^{-+}$) or scalar ($0^{++}$) mesons off a scalar target---such as 
the ${}^{4}\mathrm{He}$ nucleus~\cite{CLAS17b}---provides a fertile testing ground for exploring the hadron structure without complications from spin degrees of freedom.
In Ref.~\cite{CCLB}, we presented the most general formulation of the differential cross sections for 
the $0^{-+}$ or $0^{++}$ meson production process which involves only one or two hadronic form factors, respectively, when the target is scalar. In particular, we noted that the BSA of the coherent pseudoscalar (e.g. $\pi^0$) meson electroproduction off a scalar target (e.g. the $^4$He nucleus) should vanish as the hadronic tensor is symmetric due to a single hadronic form factor while the leptonic tensor including the electron beam polarization is totally antisymmetric. As we discussed, however, the BSA of the coherent scalar (e.g. $f_0(980)$) meson
electroproduction off the same scalar target doesn't vanish due to the two independent hadronic form factors ${\cal F}_1$ and ${\cal F}_2$ which are complex in general. These two hadronic form factors, commonly called ${\cal F}_1$ and ${\cal F}_2$ CFFs, are the main amplitudes that we analyze in the present $(3{+}1)$-dimensional scalar field model.  

In the earlier work~\cite{CCJO}, we investigated $f_{0}(980)$ production in a $(1{+}1)$-dimensional 
scalar field-theory model. In that case, however, the two CFFs ${\cal F}_1$ and ${\cal F}_2$ are not linearly
independent due to the apparent reduction of the dimensionality from $3{+}1$D to $1{+}1$D, namely, only the light-front 
plus/minus ($\pm$) components but not the transverse ($\bm\perp$) components are available. The effective reduction of
two CFFs to a single CFF renders the BSA identically zero.

Our aim in this work is to extend the previous $1{+}1$D analysis~\cite{CCJO} to the $3{+}1$D analysis, 
where both CFFs are present and can be independently extracted.
In this $(3{+}1)$D scalar field-theory framework, the hadronic current is computed to find the two CFFs, 
providing the most general Lorentz and gauge invariant formulation~\cite{CCLB}. 
Using our field theoretic model, we explicitly demonstrate that the BSA is nonzero when both CFFs are present. 
The extraction of the CFFs is made by utilizing the general
formulation of the hadronic currents presented in Ref.~\cite{CCLB}.
Indeed, the BSA can be expressed in terms of the two CFFs as
\[
\text{BSA}\ \propto\ {\cal F}_{12}^{-}\ \equiv\ {\cal F}_{1}^{*} {\cal F}_{2} - {\cal F}_{1} {\cal F}_{2}^{*}\ 
=\ 2i\,\mathrm{Im}\,({\cal F}_1^{*} {\cal F}_2).
\]
This requires that at least one of the two CFFs has nonvanishing imaginary part with a relative phase between ${\cal F}_{1}$ and ${\cal F}_{2}$.
Measuring the BSA therefore provides a clean and testable probe of the dynamics involving two CFFs.

We employ the same scalar model Lagrangian used in the previous work~\cite{CCJO}, now extended to $3{+}1$D for 
the computations of the BSA as well as the two independent CFFs. 
Our strategy is to evaluate the exact CFFs from the four-point function and 
compare them with the single-CFF description that emerges 
in the deeply virtual meson production (DVMP)~\cite{Favart:2016,Tarrach:1975,BM09,KUMER01,Bakker:2014} limit, 
where the amplitude factorizes in terms of the leading-twist GPD. This comparison is performed at generic 
momentum transfer $\Delta \equiv p - p' = q'-q$ (with $t \equiv \Delta^{2}$), including frames with a nonzero transverse 
component $\Delta_{\perp}\neq 0$, so that the extraction of the leading-twist GPD does not rely on the forward limit. 
Here, $p$ and $p'$ denote the four-momenta of the initial and
final scalar target states, while $q$ and $q'$ represent the momenta of the incoming virtual photon and the produced
meson, respectively. We identify the kinematic regions where finite $Q^{2}$ (with $Q^{2}=-q^{2}$) or beyond-leading-twist effects generate a sizable BSA and assess deviations from the DVMP-limit expectation.

In this work, we present a detailed anatomy of the contributions from each light-front (LF) time-ordered 
amplitude to the hadronic current in 3+1D. We start from the covariant four-point diagram, perform the LF time-ordered decomposition, and compute the exact two CFFs at the one loop level of the scalar field model~\cite{CCJO}. Both real and imaginary parts of the two CFFs are analyzed within the LF time-ordered framework.
These are then reduced in the DVMP limit to extract the single CFF from the leading-twist GPD. 
Here, the light-front dynamics (LFD) plays a crucial role in providing
both the skewness $\zeta (=\Delta^+/p^+)$ and the LF longitudinal momentum fraction $x$ of the parton struck by the probing virtual
photon off the target.

For the GPD analysis, we derive the analytic form of the ordinary PDF in the forward limit 
$(\zeta,t) \to 0$, as well as the analytic expression for the GPD $H(\zeta,x,t)$ at nonzero skewness $\zeta$. 
The latter is separated into contributions from the Dokshitzer-Gribov-Lipatov-Altarelli-Parisi (DGLAP)~\cite{GL72a,Dokshitzer77,AP77} region, 
$H_{\rm DGLAP} \ (\zeta \leq x \leq 1)$, and the Efremov-Radyushkin-Brodsky-Lepage (ERBL)~\cite{ER80,LB79,LB80} region, 
$H_{\rm ERBL} \ (0 \leq x \leq \zeta)$. 
We then verify the GPD sum rules through the first and second Mellin moments. 
In particular, we show that the second Mellin moment satisfies the polynomiality property in the skewness parameter 
and extract the two generalized form factors corresponding to the gravitational form factors.
The analysis carried out here is of particular relevance to the forthcoming development of the EIC and 
its program to provide the 3D tomography of the hadron structure. 

Our paper is organized as follows. In Section II, we introduce the general formulation of exclusive meson electroproduction 
and the kinematics of the VMP process off a scalar target,
$\gamma^{*} + {}^{4}\mathrm{He} \to f_{0}(980) + {}^{4}\mathrm{He}$.
Section III is devoted to the derivation of the exact form of the CFF in the VMP process within the one-loop
level of the scalar field model in (3 + 1) dimensional spacetime. Complete analyses for various LF time-ordered diagrams involved in the
VMP process are presented as well.  In Section IV, we extract
the GPD, PDF, and the EM form factor as the first Mellin moment in the DVMP limit. The polynomiality of the second Mellin moment
is explicitly shown, and the two generalized form factors corresponding to the GFFs of the scalar target are obtained.
Section~V presents our numerical results for the CFFs, BSA, PDF, GPD, and the first and second Mellin moments of the GPD. 
We summarize and conclude in Section~VI.

\section{Formulation of Exclusive Meson Electroproduction}
\subsection{Amplitude, cross section, and beam spin asymmetry}
\begin{figure}
    \centering
    \includegraphics[width=0.9\linewidth]{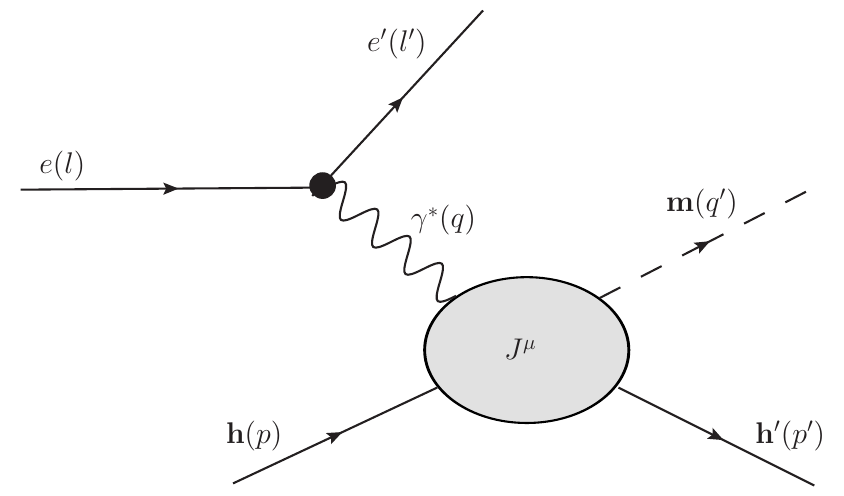}
    \caption{Momentum assignments for exclusive meson electroproduction.
    Here, $p$ and $p'$ denote the four-momenta of the initial and final scalar target states, 
while $q$ and $q'$ represent the momenta of the incoming virtual photon and the produced meson, respectively.}
    \label{fig1}
\end{figure}

We examine exclusive scalar meson electroproduction off a scalar target under the one-photon exchange mechanism,
\begin{equation}
e(l) + \mathbf{h}(p) \to e'(l') + \mathbf{m}(q') + \mathbf{h}'(p'),
\end{equation}
where the exchanged virtual photon carries momentum $q = l - l'$. Here, $\mathbf{h}(p)$ and $\mathbf{h}'(p')$ 
denote the initial and final scalar hadron (target) states with four-momenta $p$ and $p'$, respectively, 
while $\mathbf{m}(q')$ represents the produced scalar meson with four-momentum $q'$.
A schematic of the momentum flow is shown in Fig.~\ref{fig1}.

For this process, the transition amplitude takes the form
\begin{equation}
i{\mathscr M} = \frac{e^2}{q^2} L^\mu J_\mu,
\end{equation}
where we define the leptonic and hadronic currents, respectively, by
\be
L^\mu = \bar u_{e'}(l', s') \gamma^\mu u_e(l, s), 
\ee
and
\bea
J^\mu &=& \langle \mathbf{h}'(p')\, \mathbf{m}(q') 
| j^\mu(q, \Delta, \bar{p})
| \gamma^*(q)\, \mathbf{h}(p) \rangle.
\eea
Here, the hadronic current $J^\mu$ encodes all the information about the target and produced meson structure,
with $j^\mu$ containing the relevant dynamical degrees of freedom. 

Specializing to scalar ($0^{++}$) meson production off a scalar target, 
$j^\mu$ depends on three independent momenta, $q$, $\Delta = p - p'$, and $\bar{p} = p + p'$.
From the Lorentz and gauge invariance, one finds two independent form factors. 
Following~\cite{CCLB},
\bea\label{JS}
J^{\mu} &=& \left(q^{2} \Delta^{\mu}-q \cdot \Delta q^{\mu}\right) {\cal F}_{1}
+ \left(\bar{p} \cdot q \Delta^{\mu} - q \cdot \Delta \bar{p}^{\mu}\right) {\cal F}_{2}
\nonumber\\
&\equiv& A^\mu {\cal F}_1 + B^\mu {\cal F}_2.
\eea
The form factors, ${\cal F}_1$, and ${\cal F}_2$, commonly called CFFs, 
depend on the invariants $Q^2=-q^2$, $x_A \equiv \frac{Q^2}{2p\cdot q}$, and $t = \Delta^2$. 

The leptonic tensor ${\cal L}^{\mu\nu} = L^{\dagger\mu} L^\nu$ is given~\cite{CCLB} by
\be
{\cal L}^{\mu\nu}= 
q^2 \Lambda^{\mu\nu} + 2i \lambda \epsilon^{\mu\nu\alpha\beta} l_\alpha l'_\beta,
\ee
where $\Lambda^{\mu\nu} =  g^{\mu\nu} + \frac{2}{q^2} (l^\mu l'^\nu + l'^\mu l^\nu)$
denotes the symmetric leptonic tensor,
and $\lambda$ is the  electron helicity.
The hadronic tensor is ${\cal H}_{\mu\nu} = J^{\dagger}_\mu J_\nu$, so that the spin-averaged 
squared amplitude becomes
\begin{equation}\label{eq:5}
\langle |{\mathscr M}|^2 \rangle 
= \frac{e^4}{q^4} {\cal L}^{\mu\nu} {\cal H}_{\mu\nu}.
\end{equation}
This clean factorization without the Bethe-Heitler contributions allows a 
direct separation of electron kinematics from the hadronic structure, neglecting the higher-order corrections.

\begin{figure}
    \centering
    \includegraphics[width=0.95\linewidth]{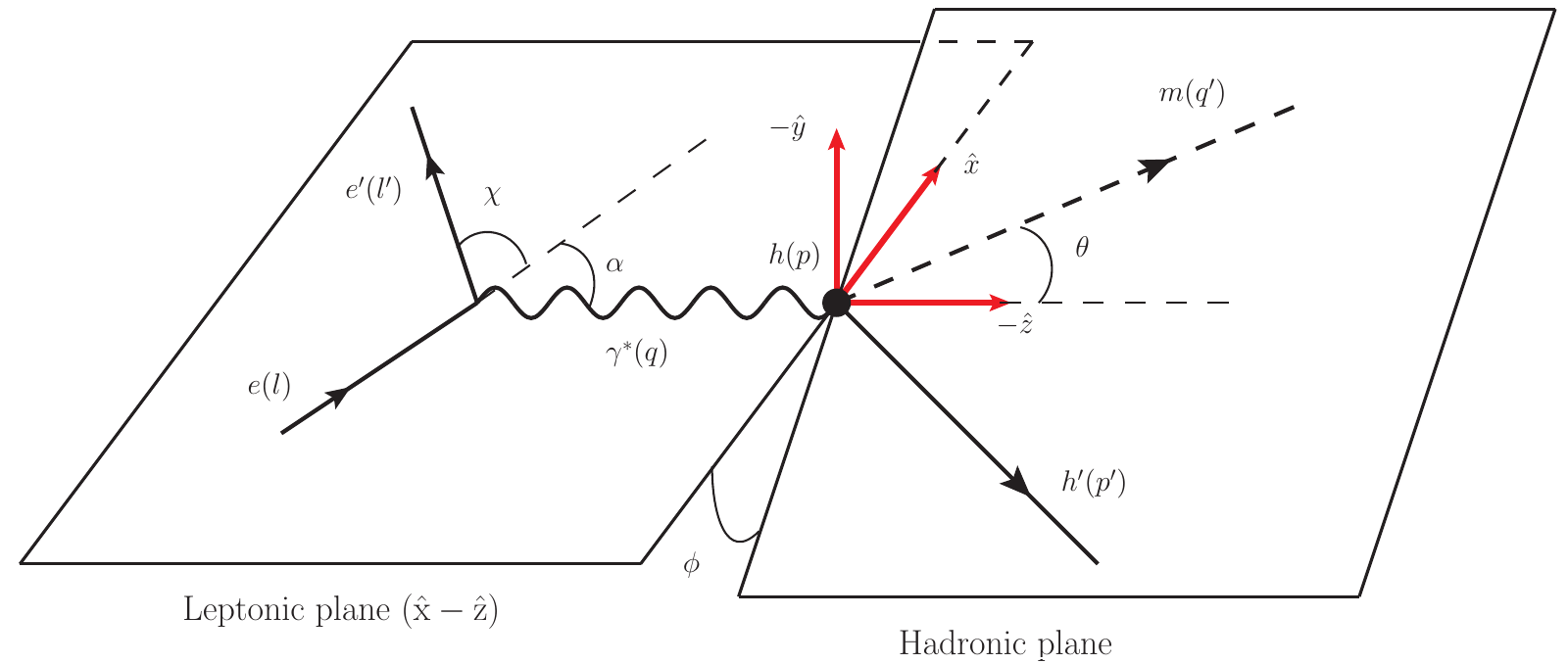}
    \caption{Schematic of the Target Rest Frame, where $\vec{\mathbf{q}} = -\left|\vec{\mathbf{q}}\right| \hat{z}$ 
    so that we have a spacelike virtual photon with $q^+ = q^0+q^3 < 0$. Note that the red arrows point in the $\hat{x}$, $-\hat{y}$, and $-\hat{z}$ directions.}
    \label{fig2}
\end{figure}

In the target rest frame (TRF), where $p^\mu=(M_T, \vec{0})$, with $M_T$ the target mass, and the $z$-axis chosen along ${\vec q}$,
the only nonvanishing spatial components of the symmetric leptonic tensor $\Lambda^{\mu\nu}$ are
\bea\label{SYT1}
  \Lambda_{xx} &=& \Lambda_T (1 + \epsilon), \; \Lambda_{yy} = \Lambda_T (1 - \epsilon), 
  \Lambda_{zz} = 2 \Lambda_T\, \epsilon_L , \nonumber\\
  \Lambda_{xz} &=& \Lambda_{zx} = - 2\Lambda_T \sqrt{\tfrac{1}{2}\,\epsilon_L(1+\epsilon)} ,
\eea
where $\Lambda_T \equiv \tfrac{1}{2} (\Lambda_{xx} + \Lambda_{yy})$ is the overall transverse strength and 
the transverse ($\epsilon$) and longitudinal ($\epsilon_L$) polarization parameters of the virtual photon are defined as
\begin{eqnarray}
  \epsilon &\equiv& \frac{\Lambda_{xx} - \Lambda_{yy}}{\Lambda_{xx} + \Lambda_{yy}} , \nonumber\\
  \epsilon_L &\equiv& \frac{\Lambda_{zz}}{\Lambda_{xx} + \Lambda_{yy}}.
\eea
Here the off–diagonal component $\Lambda_{xz}$ encodes the interference between the in–plane transverse ($x$) and longitudinal ($z$) virtual–photon polarizations, and 
 $\Lambda^{\mu\nu}H_{\mu\nu}$ reproduces the standard transverse–longitudinal interference term in the virtual–photon density–matrix formalism
 with the assignment given by Eq.~\eqref{SYT1}.

In the TRF~\cite{WJC}, as illustrated in Fig.~\ref{fig2},
the incoming and scattered electron four–momenta are chosen as
\bea\label{TRF:kk'}
&&l = E_l\bigl(1,\,  \sin\alpha,\, 0,\,- \cos\alpha\bigr),\nonumber\\
&&l' = E_{l'} \bigl(1,\, \sin(\alpha+\chi),\, 0,\, -\cos(\alpha+\chi)\bigr),
\eea
where $E_l$ and $E_{l'}=E_l -\nu$ are the incoming and scattered electron energies,
$\nu$ is the energy transfer, and $\alpha$ and $\chi$ are the scattering angles defined in Fig.~\ref{fig2}. 
In this frame the virtual photon is $q^\mu = l^\mu - l'^\mu = (\nu,0,0,|\vec q|)$, so that 
\be
E_l \sin\alpha = E_{l'} \sin(\alpha+\chi).
\ee
Combining this with
\be
Q^2 \equiv -q^2 = 4 E_l E_{l'} \sin^2\frac{\chi}{2}
\ee
one finds
\be
\sin(\alpha+\chi)\,\sin\alpha
 = \frac{Q^2}{\nu^2 + Q^2}\,\cos^2\!\frac{\chi}{2}.
\ee
Using these relations, one obtains the standard expressions for the virtual–photon polarization parameters,
\bea
\epsilon &=& \frac{1}{1 + 2\left(1 + \frac{\nu^2}{Q^2}\right)\tan^2\frac{\chi}{2}}, \nonumber\\
\epsilon_L &=& \frac{Q^2}{\nu^2}\,\epsilon.
\eea
Inserting these into the definitions of Eq.~(\ref{SYT1}), the spatial components of the symmetric leptonic tensor
$\Lambda^{\mu\nu}$ can be written explicitly in terms of $(\epsilon,\epsilon_L)$ as
\be\label{SYT2}
\Lambda_{xx} = -\frac{1+\epsilon}{1-\epsilon},\qquad
\Lambda_{yy} = -1,\qquad
\Lambda_{zz} = -2\,\frac{\epsilon}{\epsilon_L}\,\frac{\epsilon}{1-\epsilon},
\ee
with $\Lambda_{xz} = \Lambda_{zx}$ given by Eq.~(\ref{SYT1}).
With this normalization one has $g^{\mu\nu}\Lambda_{\mu\nu} = 2$.

In this frame,
the five-fold differential electroproduction cross section is given by
\begin{equation}
d\sigma 
= \frac{d^5 \sigma}{dy\, d x_A\, dt\, d\phi_{q'}\, d\phi_{l'}} 
= \kappa\, \langle |{\mathscr M}|^2 \rangle,
\end{equation}
with
\begin{equation}
\kappa = \frac{1}{(2\pi)^5}\, \frac{y\, x_A}{32\, Q^2\, \sqrt{1 + \left(\frac{2 M_T x_A}{Q}\right)^2}},
\end{equation}
where $x_A=\frac{Q^2}{2 p\cdot q}$ is the scaling variable, 
$y\equiv \frac{p\cdot q}{p\cdot l}=\frac{\nu}{E_l}=1- \frac{E_{l'}}{E_{l}}$
represents the fractional energy transfer in the TRF.

For a polarized electron beam, the squared amplitude separates into contributions 
from different photon polarizations and beam helicity~\cite{WJC},
\begin{align}\label{eq:9}
\langle |{\mathscr M}|^2 \rangle 
= \frac{e^4}{q^4} \bigg[ & 
\frac{2 q^2}{\epsilon - 1} \langle |\tau_{fi}|^2 \rangle 
+ 2 i \lambda\, \epsilon^{\mu\nu\alpha\beta} l_\alpha l'_\beta\, {\cal H}_{\mu\nu} 
\bigg],
\end{align}
where the transverse polarization can be written in terms of $y$ as 
$\epsilon =\frac{(2-y)^2}{1+(1-y)^2+2x_A^2y^2z^2}-1$ with $z^2\equiv M_T^2/Q^2$ in the TRF.
Contracting the symmetric leptonic tensor and hadronic tensor yields
\bea\label{Tfi}
\langle |\tau_{fi}|^2 \rangle 
&=& \frac12 ({\cal H}_{xx} + {\cal H}_{yy}) 
+ \frac{\epsilon}{2} ({\cal H}_{xx} - {\cal H}_{yy}) 
+ \epsilon_L {\cal H}_{zz}
\nonumber\\
&&- \sqrt{\frac12 \epsilon_L (1+\epsilon)} ({\cal H}_{xz} + {\cal H}_{zx}).
\eea

This decomposition enables the differential cross section to be expressed in a Rosenbluth-type form, 
isolating transverse (T), longitudinal (L), interference (LT), and BSA contributions. 
While the BSA term vanishes for pseudoscalar meson production due to the presence of only a single CFF,  
it does not vanish  for scalar meson production, where two independent CFFs ${\cal F}_1$ and ${\cal F}_2$ are involved. 
In this case, the BSA provides a direct probe of the imaginary parts of ${\cal F}_1$ and ${\cal F}_2$. 

The differential cross section for scalar meson production can be written
\bea
d \sigma_{\lambda} &=&  d \sigma_{T}(1+\epsilon \cos (2 \phi))+d \sigma_{L} \epsilon_{L} 
\nonumber\\
&& +d \sigma_{L T} \cos \phi \sqrt{\frac{1}{2} \epsilon_{L}(1+\epsilon)}+\lambda d \sigma_{\mathrm{BSA}}.
\eea
It should be noted that the BSA term, $d \sigma_{\mathrm{BSA}}^{\rm S}$, is proportional to 
${\cal F}_1 {\cal F}^*_2 - {\cal F}_2 {\cal F}^*_1$, as derived in~\cite{CCLB}.
The explicit expression for the BSA in coherent scalar meson electroproduction off a scalar target is given by
\bea\label{BSA_scalar}
&& \,\,\,\,
\frac{d\sigma_{\lambda=+1} - d\sigma_{\lambda=-1}}{d\sigma_{\lambda=+1} + d\sigma_{\lambda=-1}}
\nonumber\\
&&
=\frac{d \sigma_{\mathrm{BSA}}}
{d \sigma_{T} [1+\epsilon \cos (2 \phi)] + d \sigma_{L} \epsilon_{L}+d \sigma_{LT} \cos \phi \sqrt{\frac{1}{2} \epsilon_{L}(1+\epsilon)}}.
\nonumber\\
\eea
The primary focus of this model calculation is to identify kinematics that yields a sizable BSA signal. 
For further details on the construction of these expressions, we refer the reader to~\cite{CCLB}.

\subsection{Kinematics}
We start by describing the kinematics of the process in which a virtual photon scatters off a scalar target $\mathbf{h}$, 
resulting in the production of a scalar meson $\mathbf{m}$, as illustrated in Fig.~\ref{fig1}:
\begin{eqnarray}
\gamma^{*}(q)+\mathbf{h}(p) \rightarrow \mathbf{m}(q') + \mathbf{h}'(p').
\label{eq:1K}
\end{eqnarray}
To describe the momenta, we employ the light-front dynamics (LFD) variables, where a four-momentum $p$ is represented as $p=(p^+, p^-, \bm{p}_\perp)$. 
Here, $p^+ = p^0 + p^3$ denotes the longitudinal LF momentum, $p^- = p^0 - p^3$ corresponds to the LF energy, and $\bm{p}_\perp = (p^1, p^2)$ represents the transverse momentum components. We adopt the Minkowski metric convention, under which the inner product of two four-vectors is given by
$a\cdot b=\frac{1}{2} (a^+b^- + a^-b^+) - \bm{a}_\perp\cdot \bm{b}_\perp$.

Extending the previous (1+1)-dimensional analysis~\cite{CCJO} of meson production in a scalar field
model to the present (3+1)-dimensional case, we define the momenta of the target and the momentum transfer $\Delta=p-p'$ as
\begin{eqnarray}
    p &=&\Big(~ p^{+},~ \frac{M^{2}_T}{p^{+}},~ \bold{0}_{\perp} ~\Big),\nonumber\\
    \Delta &=&\Big(~\zeta p^{+},~\frac{t+ {\bm \Delta}_{\perp}^{2}}{\zeta p^{+}},~\bold{\Delta}_{\perp}~\Big),
\label{eq:2K}
\end{eqnarray}
where $\zeta=\Delta^+/p^+$ is the skewness parameter. 
The squared momentum transfer then reads
\be\label{tzeta}
t = \Delta^2 = 2 p\cdot\Delta = -\frac{\zeta^2 M^2_T + {\bm\Delta}^2_\perp}{1-\zeta}\leq 0,
\ee
which also allows us to express $\zeta$ in terms of $t$ as
\be\label{zeta}
\zeta =\frac{t + \sqrt{t^2 - 4 M^2_T (t + {\bm\Delta}^2_\perp})}{2 M^2_T},
\ee
with $0\leq\zeta\leq 1$. From Eq.~\eqref{tzeta}, one can see that the transverse component of $\Delta$ satisfies
\be
|{\bm\Delta}_\perp| = \sqrt{ (\zeta -1) t - \zeta^2 M^2_T},
\ee
where the absolute minimum $|t_{\rm min}|$ of the momentum transfer $t$ is determined by 
setting $|{\bm\Delta}_\perp|=0$ (see also Eq.~\eqref{tzeta})
for given $\zeta$ as follows
\be
|t| \geq |t_{\rm min}|=\frac{\zeta^2 M^2_T}{1-\zeta}\geq 0.
\ee
For the momenta $q'$ and $q$, where $q^2=-Q^2$ and $q^{\prime 2}=M^2_S$ with $M_S$ the mass of the produced scalar meson,
we work in the ${\bm q}_\perp=0$ frame (see Fig.~\ref{fig2}), so that ${\bm q}'_\perp={\bm\Delta}_\perp$.
In this frame,  it is convenient to parameterize $q'^-$ as being proportional to $Q^2/p^+$. 
Introducing a scale parameter $1/\zeta^{\prime}$, we write 
\be\label{eq:qprimemu}
q' = \left(\mu_S \zeta^{\prime} p^+, \frac{M^2_S + {\bm\Delta}^2_\perp}{\mu_S\zeta^{\prime} p^+}, {\bm\Delta}_\perp\right),
\ee
where $\mu_S = M_S^2/Q^2$. Then the four momentum $q=q'-\Delta$ can be written as
\be\label{eq:qmu}
q = \left(\zeta p^+ \left(\al -1 \right), q^-, {\bm 0}_\perp\right),
\ee
where $\al=\mu_S \frac{\zeta'}{\zeta}$.
Since the virtual photon carries the fixed virtuality $q^2=-Q^2$, 
$q^-=q'^- -\Delta^-$ should satisfy
\be\label{eq:qminus}
q^- = \frac{M^2_S + {\bm\Delta}^2_\perp}{\mu_S\zeta^{\prime} p^+} - \frac{t+ {\bm \Delta}_{\perp}^{2}}{\zeta p^{+}}
=\frac{Q^2}{\zeta p^+ \left(1 - \al \right)}.
\ee
Defining
$\mu'_S \equiv (M^2_S + {\bm\Delta}^2_\perp)/Q^2$ and $\tau' \equiv -(t + {\bm\Delta}^2_\perp)/Q^2$ with $\tau=-t/Q^2$, 
we obtain from Eq.~\eqref{eq:qminus} the relation between $\zeta$ and $\zeta'$:
\be\label{eq:zz'}
 \frac{\mu'_S}{\mu_S\zeta^{\prime}} + \frac{\tau'}{\zeta}
=\frac{1}{\zeta \left(1 - \al \right)}.
\ee
Solving Eq.~\eqref{eq:zz'} for the ratio $\zeta'/\zeta$, we find
\begin{equation}
    \frac{\zeta^{\prime}}{\zeta} = \frac{2\mu'_S/\mu_S}{1+\mu'_S -\tau' + \sqrt{\left(1 + \tau' + \mu'_S\right)^2 - 4\tau'}}.
    \label{eqn:zpzetaratio}
\end{equation}
We choose the solution with the $+$ sign in the denominator to ensure that in the forward limit (${\bm\Delta}_\perp \to 0$, $t \to 0$), 
the ratio $\zeta'/\zeta$ approaches unity, as required by consistency with elastic kinematics ($p^2=p^{\prime 2}=M^2_T$).
Moreover, under our chosen convention $q^+<0$,  the kinematics impose the constraint  $0\leq \mu_S\zeta' \leq \zeta \leq 1$, 
which guarantees that the virtual photon carries negative longitudinal LF momentum, consistent with our chosen frame.
In the forward limit, our expression also reproduces the result found in~\cite{CCLB} for the 1+1D case.

The scaling variable $x_A = \frac{Q^2}{2p\cdot q}$ in our chosen reference frame is given by
\be
 x_A= \frac{2(t+ {\bm\Delta}^2_\perp)} {t K_S -\sqrt{t [t-4M^2_T \left(1+{\bm\Delta}^2_\perp/t\right)](K^2_S - 4\tau') }},
\ee
where $K_S \equiv 1+\mu'_S+\tau'$.
\begin{figure}[t]
\centering
\includegraphics[width=1.0\columnwidth]{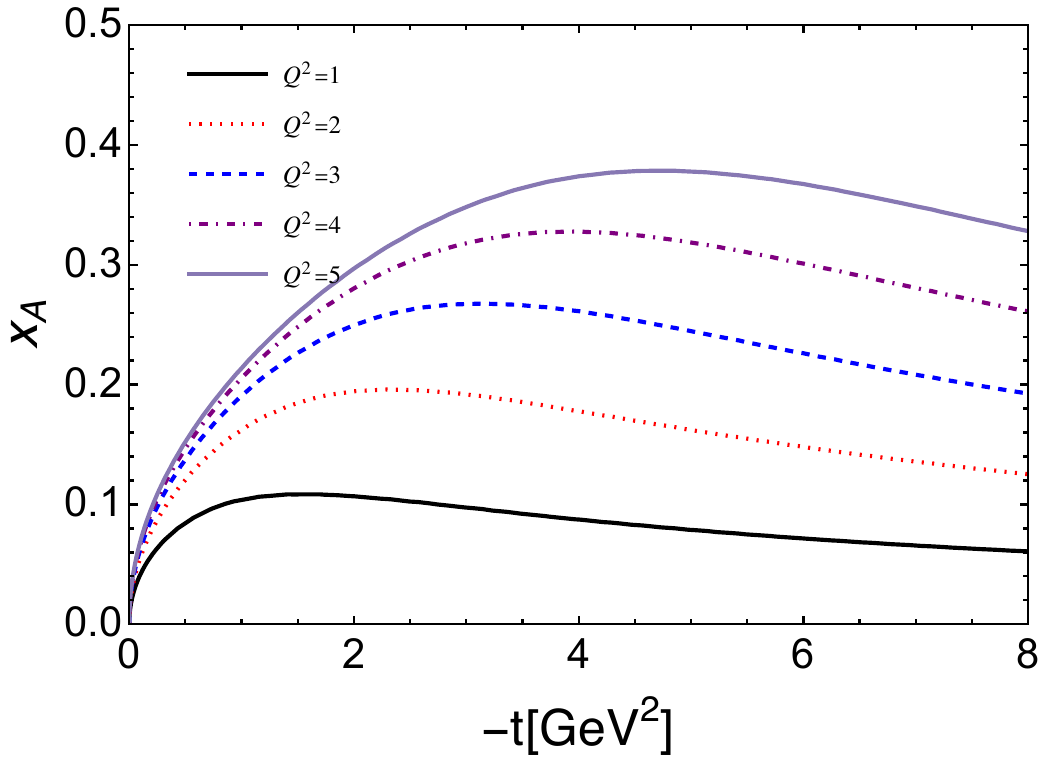}
\caption{
Scaling variable $x_A$ as a function of $-t$ 
for ${\bm\Delta}_\perp=0$ and various $Q^2$ values, 
using $(M_T, M_S) = (3.7, 0.98)$ GeV 
to model $f_0(980)$ production off a $^4{\rm He}$ target.}
\label{fig:xAt}
\end{figure}
In Fig.~\ref{fig:xAt}, we plot $x_A$ as a function of $-t$ 
for ${\bm\Delta}_\perp=0$ and various $Q^2$ values,
using the parameter sets $(M_T, M_S) = (3.7, 0.98)$ GeV 
to represent scalar meson $f_0(980)$ production off a $^4{\rm He}$ target. 
This figure illustrates the allowed ranges of $x_A$ 
for given ($-t, Q^2$) values; the lines indicate the upper bounds 
on physical values of $x_A$ for each kinematic setting.

\section{MODEL CALCULATION of Four-Point function}
In this section, we compute the matrix element of the hadronic current $J^\mu$ (see Eq.~\eqref{JS}) 
for scalar meson electroproduction off a scalar target--composed of two scalar constituents--using
a 3+1 dimensional scalar field model.

Our model Lagrangian is given by
\begin{equation}\label{eqn:lagrangian}
    \begin{split}
\mathcal{L}= & \left(\partial_{\mu} \phi_{1}+i e_{1} A_{\mu} \phi_{1}\right)^{\dagger}\left(\partial^{\mu} \phi_{1}+i e_{1} A^{\mu} \phi_{1}\right)-m_{1}^{2} \phi_{1}^{\dagger} \phi_{1} \\
& +\left(\partial_{\mu} \phi_{2}+i e_{2} A_{\mu} \phi_{2}\right)^{\dagger}\left(\partial^{\mu} \phi_{2}+i e_{2} A^{\mu} \phi_{2}\right)-m_{2}^{2} \phi_{2}^{\dagger} \phi_{2} \\
& -\frac{1}{2}\left(M^{2}_T\Phi^{2} + M_S^2 \Phi_M^2 -\partial_{\mu} \Phi \partial^{\mu} \Phi -\partial_{\mu}\Phi_M\partial^{\mu}\Phi_M\right)\\
&+g \Phi\left(\phi_{1}^{\dagger} \phi_{2}+\phi_{2}^{\dagger} \phi_{1}\right) + \Gamma\Phi_M\left(\phi_1^\dagger\phi_2 + \phi_2^\dagger\phi_1\right), 
\end{split}
\end{equation}
where $\phi_{1,2}$ are the scalar constituent fields that compose the target in our effective two-body model and carry charges $e_{1,2}$ and masses $m_{1,2}$.
In this work, we investigate the electroproduction process of a scalar meson off a scalar target in the one-loop level of the (3+1)-dimensional scalar field model extended from the conventional Wick-Cutkosky model. The same scalar field model was previously applied to the analysis of the same process in (1+1)D~\cite{CCJO}. 
In this model, the wave function is obtained as the solution of the covariant Bethe-Salpeter (BS) equation in the ladder approximation with a relativistic version of the contact interactions~\cite{Sawicki91,Sawicki92}. 
The covariant model wave function is a product of two free single-particle propagators, the Dirac delta function for the overall momentum conservation, and a constant vertex function. Consequently, all our Compton form factor calculations show various ways of evaluating the Feynman box diagrams in the scalar field model taken in the present work as well as in our previous (1+1)D work~\cite{CCJO}. While we 
simulate $\gamma^* + {}^4\mathrm{He} \to f_0(980) +  {}^4\mathrm{He}$ as an example, the internal structure of the real ${}^4\mathrm{He}$ nucleus is far more complicated, involving four nucleons (or, at a more fundamental level, twelve quarks). 
In this sense, the two scalar constituents should be understood as effective degrees of freedom that reproduce the quantum numbers of the scalar target ${}^4\mathrm{He}$.
We consider each scalar constituent carries effectively the charge of the proton and the mass of two nucleons, i.e. a pair of proton and neutron. As the average binding energy per nucleon in ${}^4\mathrm{He}$ is around 7 MeV, we take the mass of each scalar constituent as 1.865 GeV to yield the total binding energy as 30 MeV for the scalar target ${}^4\mathrm{He}$ mass 3.7 GeV. While a literal description of the actual ${}^4\mathrm{He}$ awaits for the QCD computation of nuclear systems, we take our model Lagrangian Eq.(\ref{eqn:lagrangian}) to analyze ``bare bone" structures in the one-loop amplitudes including both the ``handbag" and ``cat's ears" contributions and examine their contributions in the broad kinematics of the electroproduction of a scalar meson off a scalar target. 
Our aim in this work is to investigate the BSA in (3+1)D within the same effective model used previously in (1+1)D~\cite{CCJO}.

The field $A^\mu$ represents the virtual photon, while $\Phi$ is the scalar field describing the initial/final hadron target of mass $M_T$, and 
$\Phi_M$ denotes the electroproduced meson of mass $M_S$.
The coupling constant $g$ has units of mass and reflects the binding strength of the constituents inside the target hadron.
The parameter $\Gamma$ governs the coupling of the meson field to the scalar constituents, effectively encoding the meson-constituent vertex.
Since the meson is itself a composite particle, we can more generally model this vertex by allowing $\Gamma$ to include momentum-dependent structure, such as
$\Gamma = E_S + F_S \slashed{q}' + G_S \slashed{k} + H_S \sigma^{\mu\nu} q'_\mu k_\nu$.
While we note that additional terms can contribute due to the composite nature of the meson,
we simply take $\Gamma$ to be the identity in this work.

\subsection{Neutral Scalar Target Case}
For a neutral scalar target (total charge \(e_N=e_1 + e_2=0\)), composed of two constituents 
$Q_1$ and $Q_2$ with masses and charges \((m_1, e_1)\) and \((m_2, e_2)\) respectively, 
the covariant one-loop contributions to the hadronic current \(J^\mu\) are illustrated 
in Fig.~\ref{fig:covariantdiagrams}, where the thin and thick lines represent $\phi_1$
and $\phi_2$ fields, respectively. The diagrams correspond to
the S-channel (a), U-channel (b), and the so-called ``cat-ears" or C-channel (c) contributions.  
The Mandelstam variables $s=(p+q)^2,\; u=(p-q')^2$ are defined in the usual way,
and the inclusion of the C-channel is essential to maintain gauge invariance.

\begin{figure}
    \centering
    \includegraphics[width=1\linewidth]{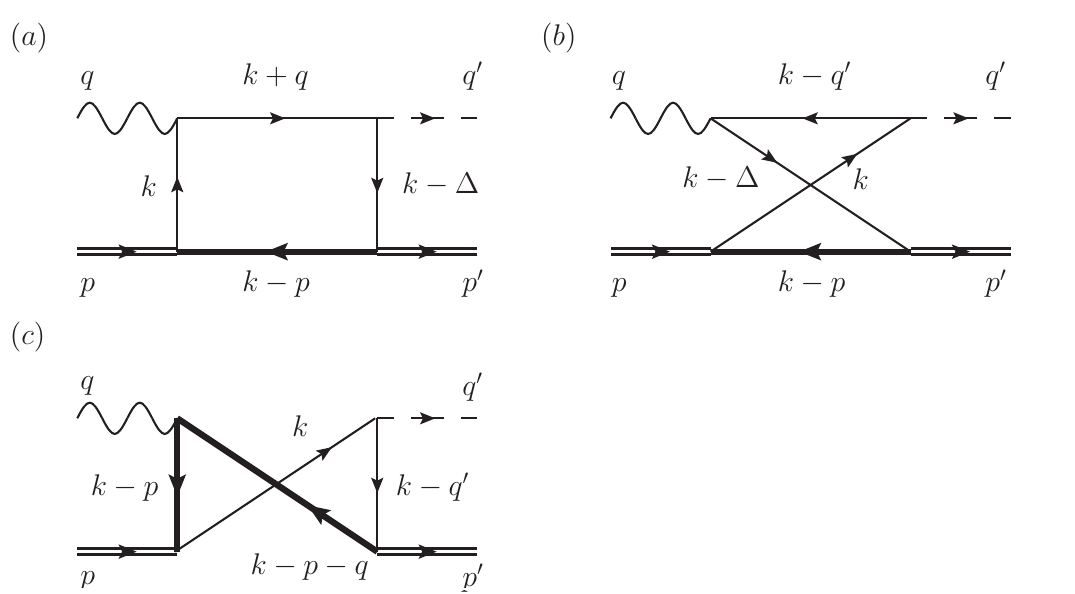}
    \caption{Relevant covariant diagrams for the reaction $\gamma^{*}(q)+\mathbf{h}(P) \rightarrow \mathbf{m}(q') + \mathbf{h}'(P')$: (a) S-Channel, (b) U-Channel, 
    and (c) ``cat-ears" diagram (denoted C-Channel).}
    \label{fig:covariantdiagrams}
\end{figure}

According to our model Lagrangian and these diagrams, the total one-loop (four-point) 
contribution to the hadronic current decomposes as
\begin{equation}
    \begin{split}
        J^\mu & = J_S^\mu + J_U^\mu + J_C^\mu,
    \end{split}
\end{equation}
with
\begin{eqnarray}\label{eq15c}
J^{\mu}_{S}&=&i e_{1} {\cal N} \int \frac{d^{4}k}{(2\pi)^{4}} 
\frac{(2 k+ q)^{\mu}}{N_{k}N_{k+q}N_{k-\Delta}D_{k-p}} ,
\nonumber\\
J^{\mu}_{U}&=&i e_{1} {\cal N}\int\frac{d^{4}k}{(2\pi)^{4}}
\frac{(2 k- 2q' +q)^{\mu}}{N_{k}N_{k-q'}N_{k-\Delta}D_{k-p}} ,
\nonumber\\
J^{\mu}_{C}&=&i e_{2} {\cal N}\int\frac{d^{4}k}{(2\pi)^{4}}
\frac{(2 k - 2p -q)^{\mu}}{N_{k}N_{k-q'}D_{k-p-q}D_{k-p}} ,
\end{eqnarray}
where the denominators arise from the intermediate scalar propagators in Fig.~\ref{fig:covariantdiagrams}.
Here,
\begin{align}
N_{p_1} &= p_1^2 - m_1^2 + i\epsilon,
\; p_1\in\{k,\;k+q,\;k-\Delta,\;k-q'\},
\nonumber\\
D_{p_2} &= p_2^2 - m_2^2 + i\epsilon,
\; p_2\in\{k-p,\;k-p-q\},
\end{align}
with $e_i$ and $m_i (i=1,2)$ denoting the charge and mass of the $i$-the constituent in the target.  
The factor \(\mathcal{N}\) contains the \(g^2\) coupling and other normalization constants.  

To proceed, we first express the covariant diagrams as an equivalent sum of their light-front (LF) time-ordered diagrams. 
The advantage of the LF calculation is that one can readily identify diagrams corresponding to valence (diagonal S-matrix elements acting on 
Fock-space states) and non-valence (off-diagonal) contributions.

In terms of the LF variables, Eq.~(\ref{eq15c}) can be 
rewritten as 
\begin{eqnarray}
    J^{\mu}_{S}&=&\frac{ie_1\mathcal{N}}{2 (2\pi)^{4}}\int \frac{dk^{+}dk^{-}d^{2}{\bm k}_{\perp}}{C_S}\nonumber\\
    &&\times\frac{(2k+q)^{\mu}}{(k^{-}-k^{-}_{i})(k^{-}-k^{-}_{f})(k^{-}-k^{-}_{t})(k^{-}-k^{-}_{b})},
    \nonumber\\
        J^{\mu}_{U}&=&\frac{ie_{1}\mathcal{N}}{2 (2\pi)^{4}}\int \frac{dk^{+}dk^{-}d^{2}{\bm k}_{\perp}}{C_U}\nonumber\\
    &&\times\frac{(2k-2q'+q)^{\mu}}{(k^{-}-k^{-}_{i})(k^{-}-k^{-}_{f})(k^{-}-k^{-}_{u})(k^{-}-k^{-}_{b})},\nonumber\\
    J^{\mu}_{C}&=&\frac{ie_{2}\mathcal{N}}{2 (2\pi)^{4}}\int \frac{dk^{+}dk^{-}d^{2}{\bm k}_{\perp}}{C_C}\nonumber\\
    &&\times\frac{(2k-2p-q)^{\mu}}{(k^{-}-k^{-}_{i})(k^{-}-k^{-}_{c})(k^{-}-k^{-}_{u})(k^{-}-k^{-}_{b})},
    \nonumber\\
\label{eq:JmuLF}
\end{eqnarray}
where $C_S = k^{+}(k^{+}+q^{+})(k^{+}-\Delta^{+})(k^{+}-p^{+})$, $C_U = k^{+}(k^{+}-q'^{+})(k^{+}-\Delta^{+})(k^{+}-p^{+})$,
and $C_C = k^{+}(k^{+}-q'^{+})(k^{+}-p^{+}-q^{+})(k^{+}-p^{+})$ are the overall LF longitudinal momentum factors for each channel, with $k^+= x p^+$.
The on-shell LF energies are given by
\begin{eqnarray}
    k^{-}_{i}&=&\frac{{\bm k}_{\perp}^{2}+m_{1}^{2}}{k^{+}}-\frac{i \epsilon}{k^{+}},\nonumber\\
    k^{-}_{f}&=&\Delta^{-}+\frac{({\bm k} -{\bm\Delta})^{2}_\perp + m_{1}^{2}}{k^{+}-\Delta^{+}}- \frac{i \epsilon}{k^{+}-\Delta^{+}},\nonumber\\
    k^{-}_{t}&=&-q^{-}+\frac{({\bm k} + {\bm q})^{2}_\perp +m_{1}^{2}}{k^{+}+q^{+}}- \frac{i \epsilon}{k^{+}+q^{+}},\nonumber\\
    k^{-}_{b}&=&p^{-}+\frac{ ({\bm k}- {\bm p})^{2}_\perp +m_2^{2}}{k^{+}-p^{+}} - \frac{i \epsilon}{k^{+}-p^{+}},\nonumber\\
    k^{-}_{u}&=&{q'}^{-}+\frac{({\bm k}-{\bm q}')^{2}_\perp +m_{1}^{2}}{k^{+}-{q'}^{+}} - \frac{i\epsilon}{k^{+}-{q'}^{+}},\nonumber\\
    k^{-}_{c}&=&p^{-}+q^{-}+\frac{ ({\bm k}- {\bm p}-{\bm q})^{2}_{\perp} +m_{2}^{2}}{k^{+}-p^{+}-q^{+}} - \frac{i\epsilon}{k^{+}-p^{+}-q^{+}}.\nonumber\\
\label{eq:onshellkminus}
\end{eqnarray}
We then integrate over the $k^-$ component, thereby placing each particle on-mass-shell with 
increasing LF time $x^+$ from left to right in the diagrammatic representation. 
This yields four LF time-ordered diagrams for each scattering amplitude, leading to a total of twelve LF time-ordered diagrams, 
as was illustrated in our previous work~\cite{CCJO} for the (1+1)-dimensional analysis. For completeness, we display them again in Fig.~\ref{fig:alldiagram}.
\begin{widetext}  
\begin{figure*}[t!]
    \centering
    \includegraphics[width=1 \textwidth]{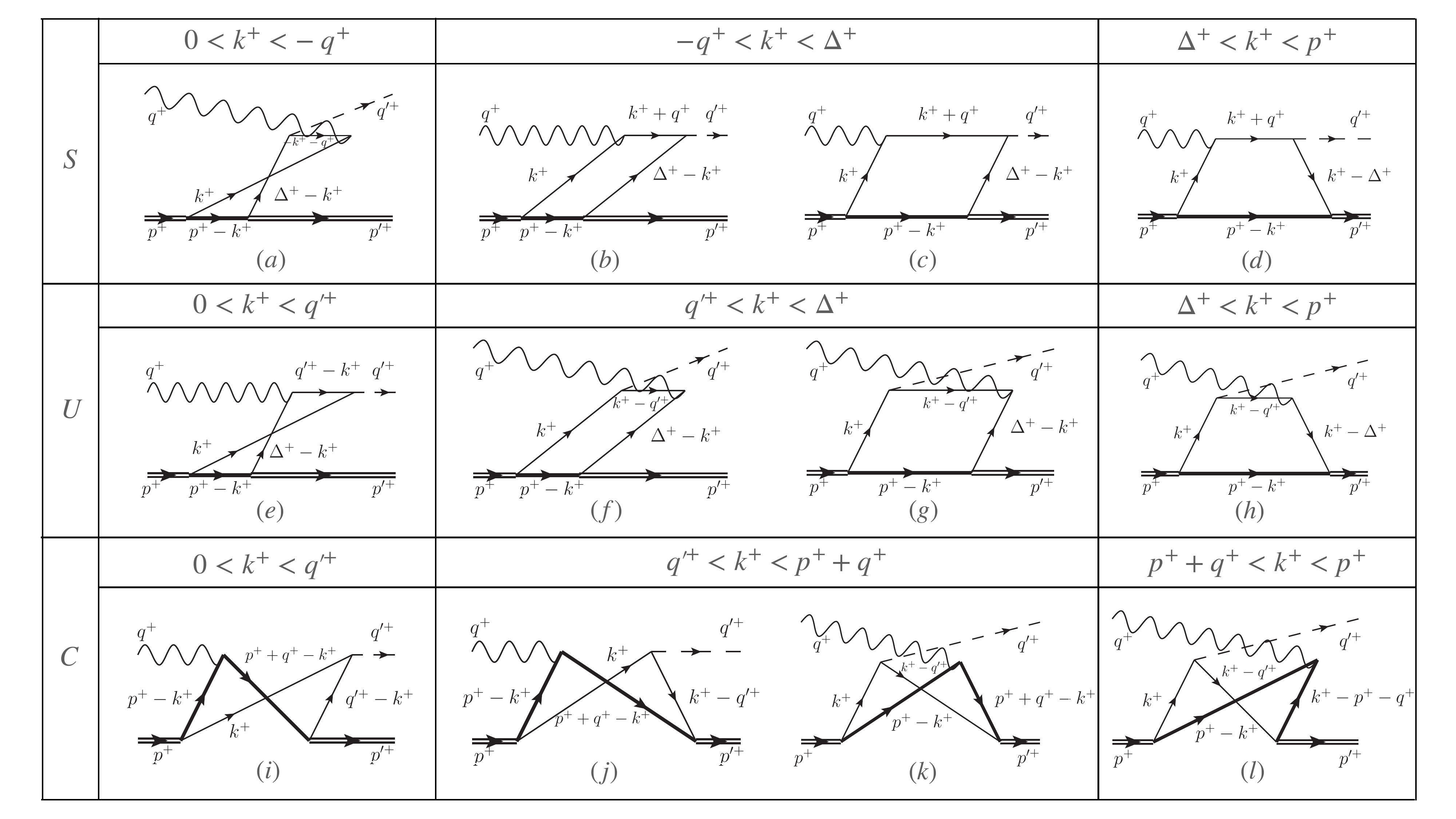}
    \caption{The LF time-ordered diagrams for the scattering amplitudues $(J^+_S, J^+_U, J^+_C)$ corresponding to (S, U, C)-channels.}
    \label{fig:alldiagram}
\end{figure*}
\end{widetext}
For the S-channel contribution $J^\mu_S$, performing the Cauchy integration over $k^-$ isolates three distinct LF time ($x^+$)-ordered residues. 
These arise from the momentum intervals $S1$ ($\Delta^+ < k^+ < p^+$), $S2$ ($-q^+ < k^+ < \Delta^+$), and $S3$ ($0 < k^+ < -q^+$), respectively.
The region $S1$ aligns with the DGLAP domain~\cite{GL72a,Dokshitzer77,AP77}, 
whereas $S2$ and $S3$ fall within the ERBL regime~\cite{ER80,LB79,LB80}.
Physically, the DGLAP region reflects valence processes that conserve particle number, while the ERBL region encompasses 
nonvalence dynamics associated with particle-number-changing transitions.

In the DGLAP kinematic domain labeled $S1$, defined by $\Delta^+ < k^+ < p^+$, 
the integral over $k^-$ in Eq.~(\ref{eq:JmuLF}) picks up its contribution from the pole at $k^-=k^-_b$, 
which lies in the upper half of the complex $k^-$ plane, while the remaining three poles are situated in the lower half.
Performing the Cauchy integration over $k^-$ in this region yields
\begin{equation}\label{eq18c}
J^{\mu}_{S,\rm hand}=\frac{-e_{1} {\cal N}}{2(2\pi)^3} \int_{\Delta^{+}}^{p^{+}} \frac{dk^{+} d^2{\bm k}_\perp}{C_S}
\frac{2 k^{\mu}_{b}+q^{\mu}}{(\Delta k^-_{bi})(\Delta k^-_{bf}) (\Delta k^-_{bt})},
\end{equation}
where we use the shorthand $\Delta k^-_{jk}= k^-_j - k^-_k$ for the energy differences. 
This term represents the so-called ``handbag'' contribution and is illustrated by the diagram in Fig.~\ref{fig:alldiagram}(d) within the S-channel.

In contrast, within the ERBL domain $S2$, where $-q^+ < k^+ < \Delta^+$, the pole structure changes: two poles, $k^-_{i}$ and $k^-_{t}$, 
are located in the lower half of the complex plane, while the other two, $k^-_{b}$ and $k^-_{f}$, lie in the upper half. 
Evaluating the integral by taking the residues at $k^-_{b}$ and $k^-_{f}$, one finds
\bea
&&\int dk^- \frac{1}{(k^- - k^-_i) (k^- - k^-_f) (k^- - k^-_t) (k^- - k^-_b)}
\nonumber\\
&&=2\pi i\left[  \frac{1}{(\Delta k^-_{bi}) (\Delta k^-_{bf}) (\Delta^-_{bt})} + \frac{1}{(\Delta k^-_{fi}) (\Delta k^-_{fb}) (\Delta^-_{ft})}
\right],
\nonumber\\
\eea
and by employing partial fraction identities such as
\bea
\frac{1}{\Delta k^-_{fi} \Delta k^-_{fb}} &=& -\frac{1}{\Delta k^-_{bi}} \left( \frac{1}{\Delta k^-_{fi}} + \frac{1}{\Delta k^-_{bf}} \right),
\nonumber\\
\frac{1}{\Delta k^-_{bf} \Delta k^-_{tf}} &=& -\frac{1}{\Delta k^-_{bt}} \left( \frac{1}{\Delta k^-_{ft}} + \frac{1}{\Delta k^-_{bf}} \right),
\eea
we identify two distinct LF time-ordered contributions in this interval:
\begin{eqnarray}
\label{eq19c}
J^{\mu}_{S, \rm stret} &=& \frac{-e_{1} {\cal N}}{2(2\pi)^3} \int^{\Delta^{+}}_{-q^{+}} \frac{dk^{+} d^2{\bm k}_\perp}{C_S}
\frac{2 k^{\mu}_{i}+q^{\mu}}{(\Delta k^-_{bi})(\Delta k^-_{fi}) (\Delta k^-_{tf})},
\nonumber\\
J^{\mu}_{S, \rm open} &=& \frac{-e_{1} {\cal N}}{2(2\pi)^3} \int^{\Delta^{+}}_{-q^{+}} \frac{dk^{+} d^2{\bm k}_\perp}{C_S}
\frac{2 k^{\mu}_{b} + q^{\mu}}{(\Delta k^-_{bi})(\Delta k^-_{bt}) (\Delta k^-_{tf})}.
\nonumber\\
\end{eqnarray}
Here, $J^{\mu}_{S, \rm stret}$ and $J^{\mu}_{S, \rm open}$ represent the ``stretched box'' and ``open diamond'' 
configurations, as depicted in Figs.~\ref{fig:alldiagram}(b) and~\ref{fig:alldiagram}(c) for the S-channel.

In the other ERBL region, denoted as $S3$ and defined by $0<k^{+}<-q^{+}$, the integral over $k^-$ in $J^{\mu}_S$ picks up its contribution 
from the pole at $k^- = k^-_i$, which is located in the lower half of the complex $k^-$ plane, while the remaining three poles reside in the upper half. 
Performing the Cauchy integration over $k^-$ in this region yields
\begin{equation}
\label{eq20c}
J^{\mu}_{S,\rm twist}=\frac{e_{1} {\cal N}}{2(2\pi)^3} \int_{0}^{-q^{+}} \frac{dk^{+} d^2{\bm k}_\perp}{C_S}
\frac{2 k^{\mu}_{i}+q^{\mu}}{(\Delta k^-_{ib})(\Delta k^-_{if}) (\Delta k^-_{it})},
\end{equation}
which corresponds to what we refer to as the ``twisted stretched box'' configuration, depicted in Fig.~\ref{fig:alldiagram}(a) for
the S-channel. 

Similarly, we can obtain the LF time-ordered amplitudes for $J^{\mu}_U$ of 
Figs.~\ref{fig:alldiagram}(e)-(h) in the U-channel and $J^{\mu}_C$ of Figs.\ref{fig:alldiagram}(i)-(l) in 
the $c$-channel.
Their explicit expressions read
\begin{eqnarray}\label{eq21c}
J^{\mu}_{U,\rm twist} &=& \frac{e_{1} {\cal N}}{2(2\pi)^3} \int_{0}^{q'^{+}} \frac{dk^{+} d^2{\bm k}_\perp}{C_U}
\frac{2 k^{\mu}_{i}-2 q'^{\mu}+q^{\mu}}{(\Delta k^-_{if})(\Delta k^-_{iu})(\Delta k^-_{ib})},
\nonumber\\
J^{\mu}_{U,\rm stret} &=& \frac{-e_{1} {\cal N}}{2(2\pi)^3} \int_{q'^{+}}^{\Delta^{+}} \frac{dk^{+} d^2{\bm k}_\perp}{C_U}
\frac{2 k^{\mu}_{i}-2 q'^{\mu}+q^{\mu}}{(\Delta k^-_{ib})(\Delta k^-_{fi})(\Delta k^-_{fu})},
\nonumber\\
J^{\mu}_{U,\rm open} &=& \frac{-e_{1} {\cal N}}{2(2\pi)^3}  \int_{q'^{+}}^{\Delta^{+}} \frac{dk^{+} d^2{\bm k}_\perp}{C_U}
\frac{-2 k^{\mu}_{b}+2 q'^{\mu}-q^{\mu}}{(\Delta k^-_{ub})(\Delta k^-_{fu})(\Delta k^-_{ib})},
\nonumber\\
J^{\mu}_{U,\rm hand} &=& \frac{-e_{1} {\cal N}}{2(2\pi)^3} \int_{\Delta^{+}}^{p^{+}} \frac{dk^{+} d^2{\bm k}_\perp}{C_U}
\frac{2 k^{\mu}_{b}-2 q'^{\mu}+q^{\mu}}{(\Delta k^-_{bf})(\Delta k^-_{bu})(\Delta k^-_{bi})}, 
\nonumber\\
\end{eqnarray}
and 
\begin{eqnarray}\label{eq22c}
J^{\mu}_{C(i)} &=& \frac{e_{2} {\cal N}}{2(2\pi)^3} \int_{0}^{q'^{+}} \frac{dk^{+} d^2{\bm k}_\perp}{C_C}
\frac{2 k^{\mu}_{i}-2 p^{\mu}-q^{\mu}}{(\Delta k^-_{ib})(\Delta k^-_{ic})(\Delta k^-_{iu}) },
\nonumber\\
J^{\mu}_{C(j)} &=& \frac{-e_{2} {\cal N}}{2(2\pi)^3} \int_{q'^{+}}^{p^{+}+q^{+}} \frac{dk^{+} d^2{\bm k}_\perp}{C_C}
\frac{-2 k^{\mu}_{i}+2 p^{\mu}+q^{\mu}}{(\Delta k^-_{bi})(\Delta k^-_{ci})(\Delta k^-_{cu})},
\nonumber\\
J^{\mu}_{C(k)} &=& \frac{-e_{2} {\cal N}}{2(2\pi)^3}\int_{q'^{+}}^{p^{+}+q^{+}} \frac{dk^{+} d^2{\bm k}_\perp}{C_C}
\frac{-2 k^{\mu}_{b}+2 p^{\mu}+q^{\mu}}{(\Delta k^-_{bi})(\Delta k^-_{bu})(\Delta k^-_{cu})},
\nonumber\\
J^{\mu}_{C(l)} &=& \frac{-e_{2} {\cal N}}{2(2\pi)^3} \int_{p^{+}+q^{+}}^{p^{+}} \frac{dk^{+} d^2{\bm k}_\perp}{C_C}
\frac{2 k^{\mu}_{b}-2 p^{\mu}-q^{\mu}}{(\Delta k^-_{bc})(\Delta k^-_{bu})(\Delta k^-_{bi})},
\nonumber\\
\end{eqnarray}
where $C_U =k^{+} (k^{+}-q'^{+}) (k^{+}-\Delta^{+}) (k^{+}-p^{+})$, $C_C = k^{+} (k^{+}-q'^{+}) 
(k^{+}-p^{+}-q^{+}) (k^{+}-p^{+})$.

\subsection{Charged Scalar Target Case}
In case of a neural target, characterized by $e_N = e_1 + e_2=0$, gauge invariance is automatically satisfied through the condition
$q \cdot J^{\mu}_{\rm tot}=0$ when the total current is given by the sum $J^{\mu}_{\rm tot}= J^{\mu}_S + J^{\mu}_U + J^{\mu}_C$. 

However, for targets carrying net charge,  such as a ``helium-like" scalar system with $e_N=-2e$, the gauge invariance requires additional contributions. 
These come in the form of so-called effectively tree (ET) level diagrams, where the external photon couples directly to the charged target line.
The ET current decomposes as
\begin{equation}
\label{eq24c}
J^{\mu}_{ET}=J^{\mu}_{S,ET} + J^{\mu}_{U,ET},
\end{equation} 
with the associated LF time-ordered diagrams illustrated in Fig.~\ref{fig:ET}.
Explicitly, the covariant expressions for these contributions are given by~\cite{CCJO}
\begin{eqnarray}
\label{eq25c}
J^{\mu}_{S, ET} &=& \frac{i e_N {\cal N}}{{\bar s}^2}
\int \frac{d^{4}k}{(2 \pi)^4}\frac{2 p^{\mu}+q^{\mu}}{N_{k} N_{k-q'} D_{k-p-q}},
\nonumber\\
J^{\mu}_{U, \rm ET} &=& \frac{i e_N {\cal N} }{{\bar u}^2}
\int \frac{d^{4}k}{(2 \pi)^4}\frac{2 p^{\mu}+q^{\mu}-2 q'^{\mu}}{N_{k} N_{k-q'} D_{k-p}}.
\nonumber \\
\end{eqnarray}
where ${\bar s}^2=s^2 -{M_{T}}^{2}$ and ${\bar u}^2=s^2 -{M_{T}}^{2}$.

\begin{figure}[t!]\centering
\includegraphics[width=0.5\textwidth]{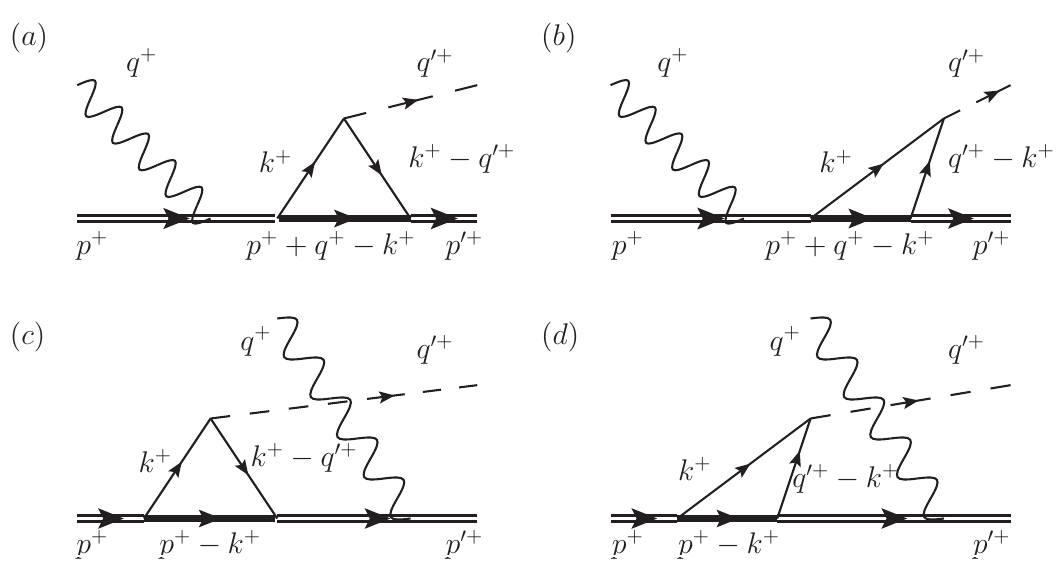}
\caption{\label{fig:ET}%
LF time-ordered effective tree diagrams in the S- and U-channels for a charged target. 
Panels (a) and (b) show the valence and nonvalence contributions to $J^{\mu}_{S, \rm ET}$, respectively, 
while (c) and (d) illustrate the corresponding contributions to $J^{\mu}_{U, \rm ET}$.
}
\end{figure}

In the LF framework, carrying out the Cauchy integration over $k^-$ in Eq.~(\ref{eq25c}) 
yields two distinct LF time-ordered contributions associated with the residue evaluations.
For the term $J^{\mu}_{S, \rm ET}$, 
these contributions emerge from two separate $k^+$ intervals:
the valence region given by $(q'^{+}<k^{+}<p^{+}+q^{+})$, depicted in Fig.~\ref{fig:ET}(a),
and the nonvalence region $(0<k^{+}<q'^{+})$, shown in Fig.~\ref{fig:ET}(b).
Similarly, for $J^{\mu}_{U, \rm ET}$, the two contributions originate from
the valence interval $(q'^{+}<k^{+}<p^{+})$ [Fig.~\ref{fig:ET}(c)] and the nonvalence interval $(0<k^{+}<q'^{+})$ [Fig.~\ref{fig:ET}(d)].
In these integrals, the poles contributing the residues differ by region:
for $J_{S,\rm ET}$, the valence term picks up the pole at $k^- = k^-_c$ (upper half),
while the nonvalence piece involves $k^- = k^-_i$ (lower half).
Similarly for $J^{\mu}_{U, \rm ET}$, residues are taken at $k^- = k^-_b$ in the valence case 
and $k^- = k^-_i$ in the nonvalence case.

Performing these contour integrations over $k^-$, we get
\begin{eqnarray}
\label{eq26c}
J^{\mu}_{S,ET, val} &=& \frac{e_N {\cal N}}{ 2(2\pi)^3 {\bar s}^2}
\int_{q'^{+}}^{p^{+}+q^{+}} \frac{dk^{+} d^2{\bm k}_\perp}{C'_S}\frac{2 p^{\mu}+q^{\mu}}{(\Delta k^-_{ic})(\Delta k^-_{cu})},
\nonumber\\
J^{\mu}_{S, ET,non} &=& \frac{e_N {\cal N}}{ 2(2\pi)^3 {\bar s}^2}
\int_{0}^{q'^{+}} \frac{dk^{+} d^2{\bm k}_\perp}{C'_S} \frac{2 p^{\mu}+q^{\mu}}{(\Delta k^-_{iu})(\Delta k^-_{ic})},
\nonumber\\
\eea
and 
\bea\label{eq:wideeq}
J^{\mu}_{U, ET, val} &=& \frac{e_N {\cal N}}{ 2(2\pi)^3 {\bar u}^2}
\int_{q'^{+}}^{p^{+}} \frac{dk^{+} d^2{\bm k}_\perp}{C'_U} \frac{2 p^{\mu}+q^{\mu}-2 q'^{\mu}}{(\Delta k^-_{ib})(\Delta k^-_{bu})},
\nonumber\\
J^{\mu}_{U, ET, non} &=& \frac{e_N {\cal N}}{ 2(2\pi)^3 {\bar u}^2}
\int_{0}^{q'^{+}} \frac{dk^{+} d^2{\bm k}_\perp}{C'_U} \frac{2 p^{\mu}+q^{\mu}-2 q'^{\mu}}{ (\Delta k^-_{iu})
(\Delta k^-_{ib})},
\nonumber\\
\end{eqnarray}
where~ $C'_S = k^{+} (k^{+}-q'^{+}) (k^{+}-p^{+}-q^{+})$ and $C'_U = k^{+} (k^{+}-q'^{+}) (k^{+}-p^{+})$.

We note that the full amplitudes also include the exchange diagrams with $Q_1 \leftrightarrow Q_2$, 
as shown in Figs.~\ref{fig:covariantdiagrams}--\ref{fig:ET}, which can be readily derived by substituting 
$Q_1 \leftrightarrow Q_2$ in the expressions above.
These contributions, along with all permutations over the distinguishable constituents $(Q_1, Q_2)$ 
and their respective masses and charges, are fully incorporated in our numerical evaluations.

Accordingly, the total matrix elements of the hadronic current $J^\mu$ for neutral and charged targets 
can be compactly written as
\bea\label{eqn:jnlo}
    J^\mu_{\rm neutral} &=& \sum_{a=S,U,C} J_a^\mu   \oplus \text{(permutations),}
    \nonumber\\
    J^\mu_{\rm charged} &=& \sum_{a=S,U,C} J_a^\mu  \oplus \sum_{a=S,U} J_{a, ET}^\mu  \oplus \text{(permutations),}
    \nonumber\\
\eea
where $J_a^\mu$ and $J_{a, ET}^\mu$ each encompass the LF time-ordered diagrams displayed 
in Figs.~\ref{fig:alldiagram} and~\ref{fig:ET}.

Once $J^\mu$ is calculated within our model, the CFFs, ${\cal F}_1$ and ${\cal F}_2$, can be extracted in a straightforward manner. 
This is done by contracting $J^\mu$ given by Eq.~\eqref{JS} with the two independent momenta $\Delta$ and ${\bar p}$, 
\bea\label{eqn:CFFsfromJ}
        \Delta\cdot J &=& (A\cdot \Delta) {\cal F}_{1}  + (B\cdot \Delta) {\cal F}_{2}, \nonumber\\
 \bar{p}\cdot J & =& (A\cdot {\bar p}) {\cal F}_{1}  + (B\cdot {\bar p}) {\cal F}_{2}.
\eea
Solving this coupled linear system for ${\cal F}_1$ and ${\cal F}_2$ is straightforward, 
as all other quantities are determined c-numbers for the chosen kinematics.
The real and imaginary parts of the CFFs follow directly.

To identify which diagrams contribute to the imaginary part of the CFFs, 
consider for example the S-channel ``handbag'' amplitude shown in Fig.~\ref{fig:alldiagram}.
An inspection of the pole structure of the energy denominators reveals that they do not all reside 
on the same side of the complex \(k^-\) plane. 
The simultaneous appearance of the \(k+q\) and \(p-k\) propagators in a light-front-time cut 
guarantees that the energy denominator \(\Delta k_{bt}^-\) becomes critical for generating an imaginary component.

In particular, for the S-channel ``handbag'' diagram, the amplitude is given by Eq.~\eqref{eq18c}, 
where the integral runs over \(\Delta^+ < k^+ < p^+\), or equivalently $\zeta < x<1$. 
The denominator \(\Delta k_{bt}^-\) is quadratic in $x$ and necessarily has a root within this interval. 
This ensures that, irrespective of the specific kinematics, the integral over this region always encounters a zero 
in the energy denominator, thereby producing an imaginary contribution to the amplitude.

In contrast, for the S-channel ``open diamond" diagram, the integral is over the range \(-q^+ < k^+ < \Delta^+\), equivalently 
$\zeta(1-\mu_s\frac{\zeta^\prime}{\zeta}) < x<\zeta$ (see Eqs.~\eqref{eq:qmu}-\eqref{eqn:zpzetaratio}). 
While $\Delta k_{bt}^-$ still has two roots for $x$, 
one of these roots always lies within $\zeta < x<1$, 
and the other may or may not fall inside $\zeta(1-\mu_s\frac{\zeta^\prime}{\zeta}) < x<\zeta$. 
Consequently, the contribution $J^\mu_{S,\rm open}$  to the imaginary part depends sensitively on the external kinematics: 
the integral over $\zeta(1-\mu_s\frac{\zeta^\prime}{\zeta})< x < \zeta $ yields an imaginary part only if this second root happens to lie within the integration bounds.

In total, there are four such diagrams -- two in the S-channel and two in the C-channel -- arising from the four-point function 
that can contribute imaginary components to the CFFs, depending on the interplay between their quadratic pole structures 
and the specific integration domains set by the kinematics.

\section{CFFs in the deeply virtual limit}
In the deeply virtual meson production (DVMP) limit, where $Q^2$ is much larger than all other relevant mass-squared scales, 
only the leading terms in $Q^2$ are retained, while contributions suppressed by powers of $1/Q^2$ are neglected.
More explicitly, in the regime \(Q^{2} \gg \left(M_{T}^{2}, M_{S}^{2}, -t\right)\), 
one has $q^- \simeq q'^- =\frac{Q^2}{\zeta p^+}$ and
the ratios $(\mu^{(\prime)}_S, \tau^{(\prime)})$ vanish. This implies $\zeta=\zeta^{\prime}$. 

In this limit, the scaling variable $x_A$ becomes \footnote{In deep inelastic scattering (DIS), 
the Bjorken variable $x_B = Q^2/(2 M_N \nu)$ is conventionally defined with respect to the nucleon mass $M_N$. 
For a target of mass $M_T$, this convention leads to the relation $M_N x_B = M_T x_A$.}
\be\label{eq:xADVMP}
x_A = \lim_{Q^{2}\rightarrow\infty}\frac{Q^{2}}{2 p\cdot q}\simeq \zeta.
\ee

\subsection{CFFs in DVMP limit}
The Lorentz factors $A^\mu$ and $B^\mu$ for the hadronic current 
$J^\mu = A^\mu {\cal F}_1 + B^\mu {\cal F}_2$ given by Eq.~\eqref{JS} reduce in this DVMP limit to
\bea\label{eq:ABDVMP}
A^\mu &=& (\Delta \cdot q) q^{\mu}-q^{2} \Delta^{\mu} \nonumber\\
&\simeq& \frac{Q^2}{2} (2 q^{\prime \mu} - q^\mu)\equiv A^\mu_{\rm DVMP},
\nonumber\\
B^\mu &=& (\Delta \cdot q)\bar{p}^\mu - (\bar{p}\cdot q)\Delta^\mu 
\nonumber\\
&\simeq& - \frac{Q^2}{\zeta}(\Delta^\mu -\zeta p^\mu)\equiv B^\mu_{\rm DVMP}.
\nonumber\\
\eea
Here, we have used $\Delta\cdot q \simeq Q^2/2$ and $\bar{p}\cdot q \simeq (2-\zeta)Q^2/(2\zeta)$, 
as obtained from Eq.\eqref{eq:2K} to leading order in $Q^2$.

Thus, the scattering amplitude in the DVMP limit may be written as
\be\label{eq:JDVMP}
J^\mu_{\rm DVMP}= A^\mu_{\rm DVMP} {\cal F}^{\rm DVMP}_1 +  B^\mu_{\rm DVMP} {\cal F}^{\rm DVMP}_2.
\ee
Taking this limit greatly reduces the number of non-vanishing NLO diagrams; the four-point function of the surviving diagrams reduces
to a three-point function via the operator product expansion of the \(\mathcal{O}(Q^2)\) propagators. 

In the leading-twist approximation, 
the amplitude $J^\mu_{\rm DVMP}$ that predominantly contributes to meson production in the deeply virtual region
arises from the S-and U-channels, i.e.,
\be\label{eq:JDVMPSU}
J^\mu_{\rm DVMP}= J^\mu_{S,\rm DVMP} + J^\mu_{U,\rm DVMP}.
\ee
The propagators corresponding to the hard scattering subprocess 
in the S- and U-channels (see Eq.~\eqref{eq15c}) simplify in the DVMP limit:
\bea
    \frac{2k^{\mu}+q^{\mu}}{(k+q)^{2}-m^{2}_1 + i\ep}&\simeq&\Big(\frac{\zeta}{x-\zeta + i\ep}\Big)\frac{(2k+q' -\Delta)^{\mu}}{Q^{2}},\nonumber\\
    \frac{2k^{\mu}-2{q'}^{\mu} + q^{\mu}}{(k-q')^{2}-m^{2}_1 + i\ep}&\simeq&\Big(\frac{-\zeta}{x- i\ep}\Big)\frac{(2k- q' - \Delta)^{\mu}}{Q^{2}},
\eea
where we have used the kinematic identities $2k + q= 2k + q' -\Delta$ and $2k - 2q' + q= 2k -q' -\Delta$.
Starting from Eq.~\eqref{eq15c}, we thus obtain
\begin{widetext}
\begin{eqnarray}\label{eq15cDVMP}
J^{\mu}_{\rm DVMP}&=&i e_{1} {\cal N}\frac{\zeta}{Q^2} 
\int \frac{d^{4}k}{(2\pi)^{4}} \left( \frac{(2k+q'-\Delta)^{\mu}}{x-\zeta + i\ep} - \frac{(2k- q' - \Delta)^{\mu}}{x - i\ep}\right)
 \frac{1}{N_{k}N_{k-\Delta}D_{k-p}},
 \nonumber\\
&=&i e_{1} {\cal N}\frac{\zeta}{Q^2} 
\int \frac{d^{4}k}{(2\pi)^{4}} \left( \frac{1}{x-\zeta + i\ep} - \frac{1}{x - i\ep}\right)
 \frac{  S^\mu }{N_{k}N_{k-\Delta}D_{k-p}},
\end{eqnarray}
\end{widetext}
where we have defined $S^\mu=(2k-\Delta)^\mu + \frac{(2x-\zeta)}{\zeta} q^{\prime\mu}$. 
For the LF components, $\mu =+$ and $-$, we find $S^+= (2x-\zeta)p^+$ and 
$S^- \simeq \frac{(2x -\zeta)Q^2}{\zeta^2 p^+}=\frac{Q^2}{(\zeta p^+)^2}S^+$ in the DVMP limit.
Comparing with Eqs.~\eqref{eq:ABDVMP} and~\eqref{eq:JDVMP} together with Eq.~\eqref{eq15cDVMP}, 
we obtain
\bea
J^+_{\rm DVMP} &=& \frac{\zeta Q^2 p^+}{2} {\cal F}^{\rm DVMP}_1,
\label{eq:JplusDVMP}\\[1ex]
J^-_{\rm DVMP} &\simeq& \frac{Q^2}{ (\zeta p^+)^2} J^+_{\rm DVMP} \nonumber\\
&=& \frac{Q^4}{2\zeta p^+}\left[ {\cal F}_1^{\rm DVMP} + \frac{2\zeta M^2_T}{Q^2} {\cal F}_2^{\rm DVMP} \right].
\label{eq:JminusDVMP}
\eea
Solving Eqs.~\eqref{eq:JplusDVMP} and~\eqref{eq:JminusDVMP}, we find

\begin{align}\label{FCCDVMP}
{\cal F}^{\rm DVMP}_1 
&= \frac{2}{\zeta Q^2 p^+} J^+_{\rm DVMP} \nonumber\\
&= i e_{1} {\cal N} \frac{2 \zeta^2}{Q^4}
\int \frac{d^{4}k}{(2\pi)^{4}} 
\left(
\frac{1}{x - \zeta + i\epsilon}
- \frac{1}{x - i\epsilon}
\right)
\nonumber\\
&\phantom{= i e_{1} {\cal N} \frac{2 \zeta^2}{Q^4}
\int \frac{d^{4}k}{(2\pi)^{4}} }
\times \frac{(2x - \zeta)}{N_{k} N_{k - \Delta} D_{k - p}},
\end{align}
and ${\cal F}_2^{\rm DVMP}=0$ in this DVMP limit.
Thus, for the $\mu=\pm$ components in the DVMP limit, we arrive at a single CFF, 
obtained in a manner that is independent of the specific current 
component.\footnote{While we choose a reference frame such that $Q^2$ flows entirely along the longitudinal direction, i.e., ${\bm q}_\perp=0$,
one can likewise show that in a more general frame where both $q^-$ and ${\bm q}_\perp$ are nonzero,
our conclusion remains valid when derived from the \(\mu = \perp\) component.}

\begin{figure}[t]\centering
\includegraphics[width=\linewidth]{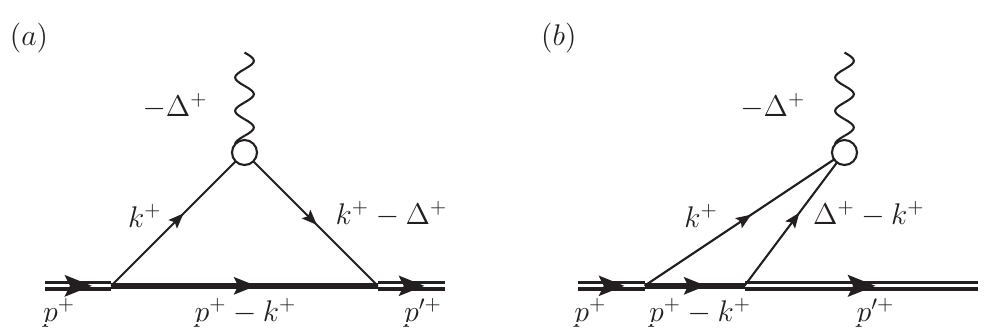}
\caption{\label{fig:trianglediagrams}%
Diagrams for GPDs in the DVMP limit.
The sum $J^{+\rm DVMP}_{S,\rm hand} + J^{+\rm DVMP}_{U,\rm hand}$
gives the valence contribution (a) in the DGLAP ($\zeta\leq x\leq 1$) region, while  
$J^{+\rm DVMP}_{S, \rm twist} + J^{+\rm DVMP}_{U,\rm stret}$ gives 
the nonvalence contribution (b) in the ERBL ($0\leq x \leq \zeta$) region. 
The small white blob denotes the nonlocal constituent-gauge-boson vertex.
}
\end{figure}

In terms of LF variables, the covariant integral for the three-point function in Eq.~\eqref{FCCDVMP}
\be\label{Threepoint}
I_3 \equiv \int \frac{d^{4}k}{(2\pi)^{4}} \frac{1}{N_{k}N_{k-\Delta}D_{k-p}} = I_{\rm DGLAP} + I_{\rm ERBL}
\ee
yields two nonzero contributions from the Cauchy integration over $k^-$:
(I) for $0< k^+< \Delta^+$ (ERBL region), with the residue at the pole $k^-=k^-_i$ in the lower half-plane, and 
(II) for $\Delta^+ < k^+ <p^+$ (DGLAP region), with the residue at $k^-=k^-_b$ in the upper half-plane.

Explicitly, we obtain
\begin{eqnarray}    
    I_{\rm DGLAP} (k^-=k^-_b) &=&i \int_{\Delta^+}^{p^+} \frac{dk^{+} d^{2}{\bm k}_{\perp}}{2(2\pi)^{3}}\frac{1}{C_3 (\Delta k^-_{bi})(\Delta k^-_{bf})},\nonumber\\
    I_{\rm ERBL} (k^-=k^-_i) &=&-i \int_{0}^{\Delta^+} \frac{dk^{+} d^{2}{\bm k}_{\perp}}{2(2\pi)^{3}}\frac{1}{C_3 (\Delta k^-_{ib})(\Delta k^-_{if})},
    \nonumber\\
\end{eqnarray}
where $C_3 = k^{+}(k^{+}-\Delta^{+})(k^{+}-p^{+})$.

In this way, we obtain the factorized form of the single CFF ${\cal F}^{\rm DVMP}_1$ 
and the corresponding DVMP amplitude $J^+_{\rm DVMP}$ in the DVMP limit. 
Hence, ${\cal F}^{\rm DVMP}_1$ is directly related to the leading-twist GPD, 
as we discuss in the next subsection.

\subsection{Leading twist GPD}
As discussed above, the amplitude $J^+_{\rm DVMP}$ factorizes into hard and soft parts, 
allowing ${\cal F}^{\rm DVMP}_1$ to be directly related to the leading-twist GPD.
In this context, it is instructive to compare with the reduced three-point function 
associated with the electromagnetic (EM) form factor $F_{\rm EM}(t)$ of the charged target, 
defined via $J^{\mu}_{\rm EM} = (p+p')^\mu F_{\rm EM}(t=-{\bm\Delta}^2)$, where the EM current is given by
\begin{eqnarray}\label{JEM}
J^{\mu}_{\rm EM}&=&i {\cal N}
\int \frac{d^{4}k}{(2\pi)^{4}} \frac{\delta(x-k^+/p^+)(2k-\Delta)^\mu}{N_{k}N_{k-\Delta}D_{k-p}},
\end{eqnarray}
where $\delta(x-k^+/p^+)$ represent the nonlocality of the struck constituent-gauge boson vertex shown in Fig.~\ref{fig:trianglediagrams}.

Since $J^{\mu}_{\rm DVMP}\propto J^{\mu}_{\rm EM}$, the leading-twist GPD can be extracted from the plus component of the current, i.e., $(2k -\Delta)^+=(2x-\zeta)p^+$, and is thus related to the CFF ${\cal F}_1$.
Explicitly, the leading-twist DVMP amplitude takes the factorized form
\be\label{Tw2DVMP}
J^+_{\rm DVMP} =\frac{ e_1 \zeta}{Q^2} 
\int dx \left( \frac{1}{x-\zeta + i\ep} - \frac{1}{x - i\ep}\right) H(x,\zeta,t),
\ee
where
\be\label{HGPD}
H(x,\zeta,t)=i {\cal N}p^+\int\frac{dk^- d^2{\bm k}_\perp}{2(2\pi)^4} \frac{(2k-\Delta)^+}{N_{k}N_{k-\Delta}D_{k-p}}
\ee
defines the leading-twist GPD.
After the integration over $k^-$, the current can be written as the sum of the two LF time-ordered diagrams (see Fig.~\ref{fig:trianglediagrams}),
i.e., 
\begin{equation}
\label{eq32c}
H=
\begin{cases}
~H_{\rm ERBL}(x,\zeta,t) & \text{for }~ 0\leq x\leq \zeta, \\
~H_{\rm DGLAP}(x,\zeta,t) & \text{for }~ \zeta\leq x\leq 1,
\end{cases}
\end{equation}
where
\begin{eqnarray}
    H_{\rm DGLAP}=\frac{-\mathcal{N}(2x-\zeta)}{x(x-\zeta)(x-1)p^+}\int \frac{d^{2}{\bm k}_{\perp}}{2 (2\pi)^3}\frac{1}{(\Delta k^-_{bi})(\Delta k^-_{bf})},\nonumber\\
    H_{\rm ERBL}=\frac{\mathcal{N}(2x-\zeta)}{x(x-\zeta)(x-1)p^+}\int \frac{d^{2}{\bm k}_{\perp}}{2 (2\pi)^3}\frac{1}{(\Delta k^-_{if})(\Delta k^-_{ib})}.\nonumber\\
\end{eqnarray}

The analytic forms of the GPDs $H(x,\zeta,t)$ in the equal-mass case ($m_1 = m_2 = m$) are given by
\bea\label{eq:GPDnonperp}
    H_{\rm DGLAP}
    &=& \frac{ {\cal N} }{2 (2\pi)^2} \frac{(1-x) (2x -\zeta)/(1-\zeta)}{\sqrt{(a_D -b_D)^2 + a_D c^2_D}}
\nonumber\\
&&\times\biggl\{  - \tanh^{-1}\biggl(\frac{a_D - b_D}{ \sqrt{ (a_D - b_D)^2 + a_D c^2_D}} \biggr)
\nonumber\\
&&+ \tanh^{-1}\biggl(\frac{2 (a_D - b_D) + c_D^2}{ 2\sqrt{ (a_D - b_D)^2 + a_D c^2_D}} \biggr)
\biggr\},
\nonumber\\
H_{\rm ERBL} 
&=& \frac{ {\cal N} }{2 (2\pi)^2} \frac{x (2x -\zeta)/\zeta}{ \sqrt{(a_E -b_E)^2 + a_E c^2_E}}
\nonumber\\
&&\times\biggl\{  - \tanh^{-1}\biggl(\frac{a_E - b_E}{ \sqrt{ (a_E - b_E)^2 + a_E c^2_E}} \biggr)
\nonumber\\
&&+ \tanh^{-1}\biggl(\frac{2 (a_E - b_E) + c_E^2}{ 2\sqrt{ (a_E - b_E)^2 + a_E c^2_E}} \biggr)
\biggr\},
\eea
where
\bea
a_D &=& a_E = m^2 - x(1-x)M^2_T, \nonumber\\
b_D &=& m^2 
+ \frac{(1-x)\left[ \zeta^{2} M_{T}^{2} - \zeta ( x M_{T}^{2}  + t) + x\left( {\bm\Delta}_{\perp}^{2} + t \right)\right]}{\zeta (1-\zeta)}, 
\nonumber\\
b_E &=&  m^2 
+ \frac{x\left[ x({\bm\Delta}^2_\perp + t) -\zeta t\right]}{\zeta^2}, \nonumber\\
c_D &=&  2 \frac{1-x}{1-\zeta} |{\bm\Delta}_\perp|,\;
c_E = 2 \frac{x}{\zeta}|{\bm\Delta}_\perp|.
\eea
We should note that $|{\bm\Delta}_\perp|=\sqrt{(\zeta -1)t- \zeta^2 M^2_T}$.

If we take the reference frame where ${\bm\Delta_\perp}=0$, then $\zeta$ and $t$ are not independent, 
but connected through $-t=|t_{\rm min}|=\zeta^2 M^2_T/(1-\zeta)$. 
In this case, the GPD can be expressed in terms of $H(x,\zeta(t_{\rm min}))$, yielding
\bea\label{eq:ERBLzeta}
H_{\rm DGLAP} (x, \zeta(t_{\rm min}))&=& 
\frac{{\cal N} (\zeta-1)(2x-\zeta) \log\! A_{\rm D}}
{8\pi\zeta M^2_T [(\zeta -2)x +1]},
\nonumber\\
H_{\rm ERBL} (x, \zeta(t_{\rm min})) &=& 
\frac{{\cal N} (\zeta-1)(2x-\zeta) \log\! A_{\rm E}}
{8\pi\zeta M^2_T [(\zeta -2)x +1]},
\nonumber\\
\eea
where
\bea\label{eq:ADAE}
A_{\rm D} &=& \frac{(\zeta-1)^2(m^2 - M^2_T x(1-x))}{(\zeta-1)^2 m^2 + M^2_T (x-1) (x-\zeta)},
\nonumber\\
A_{\rm E} &=&\frac{(\zeta-1)(m^2 - M^2_T x(1-x))}{(\zeta-1)m^2 + M^2_T x (x-\zeta)},
\eea
and the range $0\leq \zeta(t_{\rm min})\leq 1$ corresponds to $0\leq |t_{\rm min}| < \infty$.

\subsection{PDF and Lowest Moment of GPD}
In the limit $(\zeta,t)\to 0$, we recover the ordinary parton distribution function (PDF) $q(x)$, i.e., 
$H(x,0,0)=H_{\rm DGLAP}(x,0,0)=q(x)$.
For the case of equal constituent masses, $m_1=m_2=m$, we obtain the analytic form 
\be
q(x) = \frac{{\cal N}}{2(2\pi)^2}\frac{x(1-x)}{x(1-x)M^2_T -m^2}.
\ee
The normalization constant ${\cal N}$ is fixed by 
\be
\int^1_0 dx\; q(x)=1, 
\ee
which imposes unit probability to find the scalar constituent in the target within this two-body model.
Since the two scalar fields $\phi_1$ and $\phi_2$ have equal mass and share the total momentum symmetrically,
the first moment satisfies
$\int_0^1 dx \; x\, q(x) = 1/2$
for each constituent, and therefore the total momentum fraction carried by the two constituents fulfills
$\sum_{i=1}^2 \int_0^1 dx \; x\, q_i(x) = 1$.

The same normalization constant ${\cal N}$ is consistently used in the analyses of the CFFs and BSA.
For comparison, the corresponding PDF in (1+1) dimensions is given by~\cite{CCJO}
\be
q(x) \propto \frac{x(1-x)}{ [x(1-x)M^2_T -m^2]^2}.
\ee

It is also well known that the lowest moment, i.e. the first Mellin moment, of the leading-twist GPD provides
the electromagnetic form factor of the target meson as given by
\begin{equation}
    F_{\text{EM}}\left(t\right) = \int^1_0 \frac{d x}{1-\zeta/2} H\left(x, \zeta, t\right).
\end{equation}
This well-known GPD sum-rule for the total result of summing the DGLAP and ERBL regions holds for all components of the hadronic current as one may understand from $J^{\mu}_{\rm DVMP}\propto J^{\mu}_{\rm EM}$. However, one should note that the respective correspondence of the DGLAP and ERBL regions to the valence and non-valence parts of the electromagnetic form factor holds only for the light-front plus component of the hadronic current but not for any other components of the hadronic current. This point was already addressed in our previous work~\cite{CCJO}, noting that the decomposition of the form factor between the valence and nonvalence contributions depends on which component of the current is used for the calculation. Therefore, the correspondence between the DGLAP vs. ERBL regions and the valence vs. non-valence parts of the electromagnetic form factor should be understood with great care in taking the component of the hadronic current.

\subsection{Polynomiality of Higher Moments}

In discussing the polynomiality of the second and higher Mellin moments of GPDs, it is advantageous to 
introduce the symmetric variables $(X, \xi)$ rather than $(x, \zeta)$. This notation renders the underlying 
kinematic relations more transparent and facilitates a direct comparison with the conventional form of the polynomiality condition.

For this analysis, we follow the symmetric notation for the GPDs~\cite{Ji_1998,PW99,PS18,BR08}
\be\label{eq:SymX}
\xi = \frac{\zeta}{2-\zeta},\;  X= \frac{x - \zeta/2}{1-\zeta/2},
\ee
where $0\leq \xi\leq 1$ and the support is $-1\leq X\leq 1$. The  GPD corresponding to $H(x, \zeta, t)$ is then defined as
\be\label{eq:GPDsym}
H (X, \xi, t) = H \left(\frac{\xi + X}{\xi +1},\frac{2\xi}{\xi+1},t \right).
\ee
Reflection about $X=0$ yields the symmetric relation $H(X,\xi, t)=H(-X, \xi, t)$.

The polynomiality conditions~\cite{Ji_1998,PW99,PS18,BR08,Son25} follow from the Lorentz invariance, time reversal, and Hermiticity. 
In general, the polynomiality conditions for the moments of the GPDs, defined by
\be\label{eq:Poly1}
F^{(n)}(\xi,t)=\int^1_{-1} dX  X^{n-1} H(X,\xi,t)=\sum_{m=0,{\rm even}}^n \xi^m A_{nm}(t),
\ee
require that the highest power of $\xi$ in the polynomial expression of $F^{(n)}(\xi,t)$ does not exceed $n$~\cite{JiMel97,RAD99,CJL01},
where $A_{nm}(t)$ are the generalized form factors.
The first ($n=1$) moment corresponds to the EM form factor, i.e. $F^{(1)}(\xi,t)=F_{\rm EM}(-t)$, which is independent $\xi$.
The second ($n=2$) moment satisfies the $\xi$-dependent polynomiality relation
\be\label{eq:Poly2}
F^{(2)}(\xi, t) = A_{20}(t) + \xi^{2} A_{22}(t),
\ee
where 
$A_{20}(t) = F^{(2)}(0,t)$ and $A_{22}(t)$ correspond to the gravitational form factors (GFFs) of the scalar target.  
In the following section, we explicitly demonstrate that our model satisfies the polynomiality condition of Eq.~\eqref{eq:Poly2}.

\section{Numerical Results}
In this numerical section, we first fix the normalization constant ${\cal N}$ from the ordinary PDF $q(x)$ and then use the same value of ${\cal N}$ for all other physical quantities.
In our numerical analysis, we take $m_1=m_2=m=1.865$ GeV and $M_T=3.7$ GeV for the scalar He-like target,
but the method can be readily generalized to unequal constituent masses.
This choice corresponds to a weakly bound two-body system with binding energy 
$\epsilon_{\rm BE} =  (m_1 + m_2) - M_T= 30~\mathrm{MeV}$ (corresponding to about 7 MeV per nucleon)
and serves as a He-like scalar target in our effective model framework.

\subsection{PDF, GPD, and Mellin moments}
\begin{figure}
    \includegraphics[width=\linewidth]{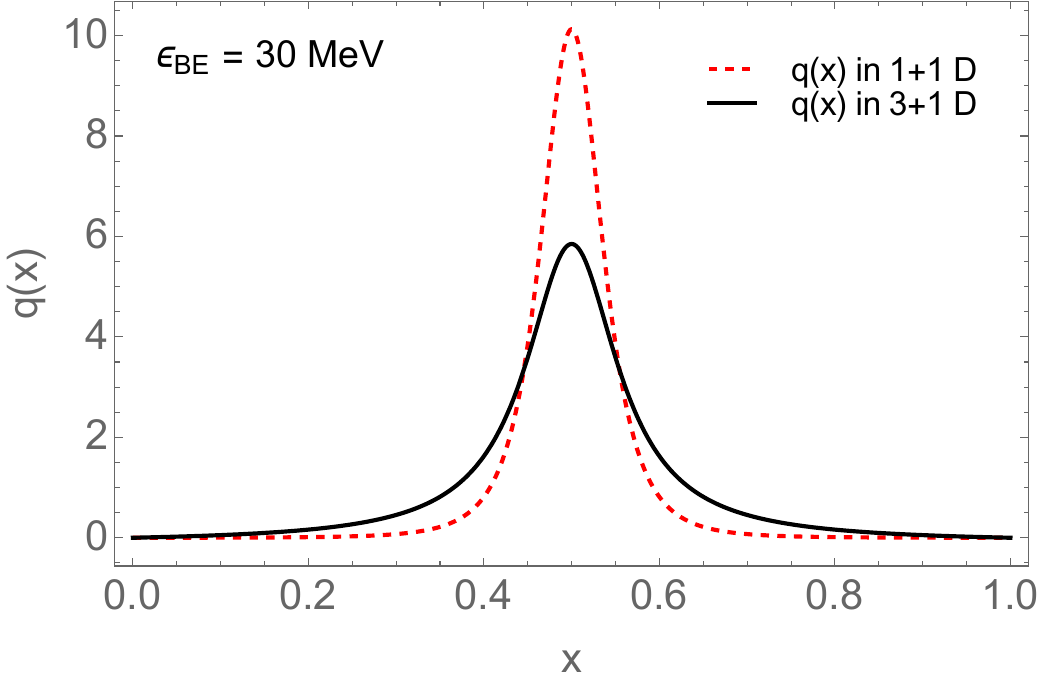}
    \caption{Ordinary PDF $q(x)$ of a He-like target with mass $M_T=3.7$ GeV 
    and binding energy $\ep_{\rm EB}=30$ MeV (i.e. $m_1 = m_2 =m= 1.865$ GeV).}
    \label{fig:PDF}
\end{figure}
\begin{figure}
  \includegraphics[width=\linewidth]{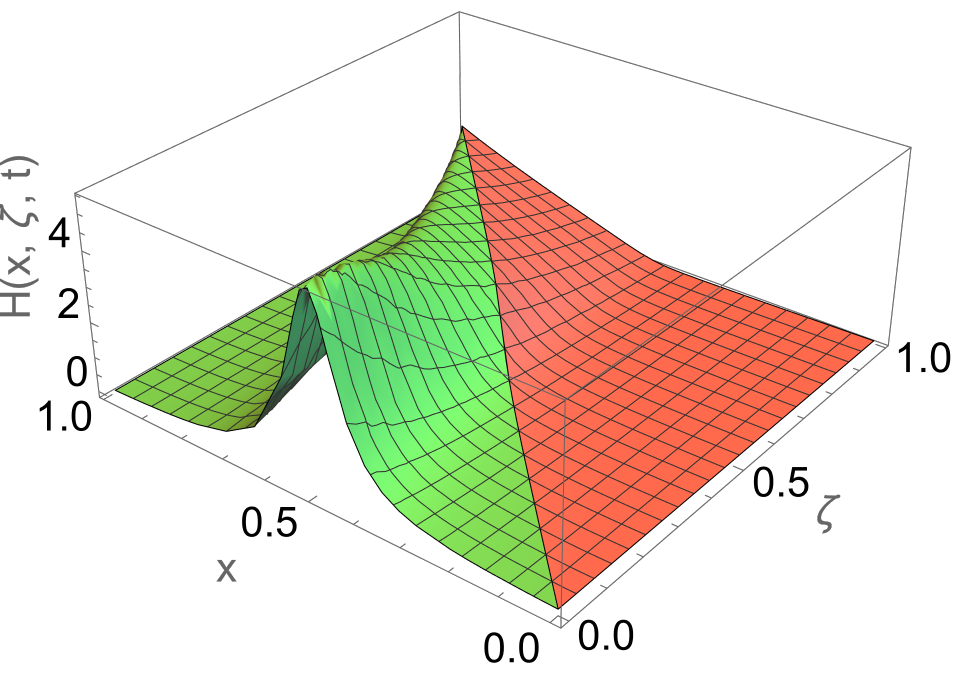}
  \includegraphics[width=\linewidth]{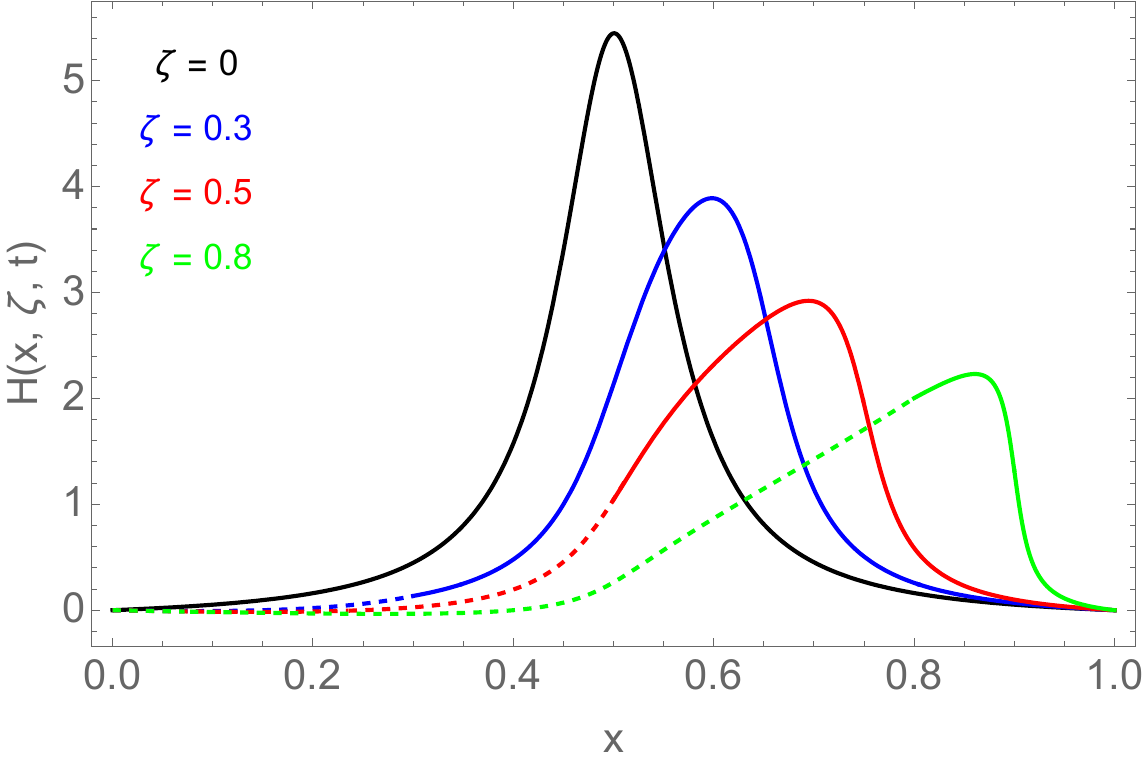}
   \caption{Three-dimensional (3D, top) and  two-dimensional (2D, bottom) GPDs at fixed $-t=0.1$ GeV$^2$ 
obtained from Eq.~\eqref{eq:GPDnonperp}, i.e., in the ${\bm\Delta}_\perp\neq 0$ frame. }
  \label{fig:GPDHe}
\end{figure}
\begin{figure}
  \includegraphics[width=\linewidth]{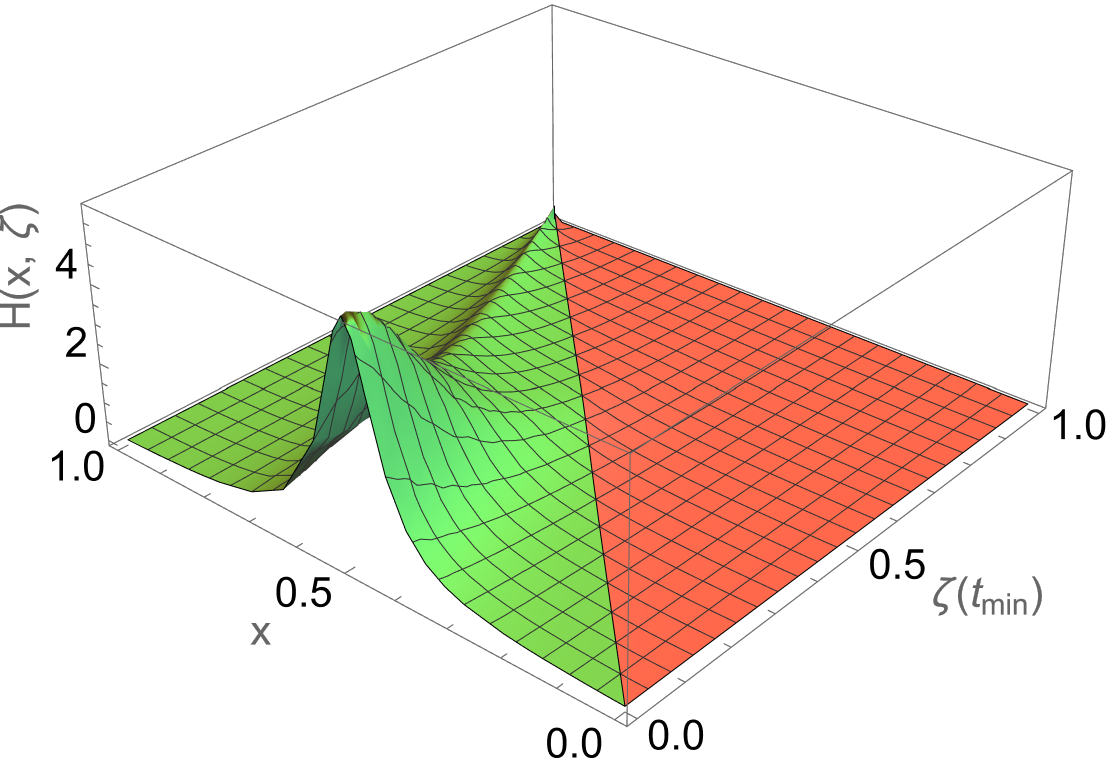}
  \includegraphics[width=\linewidth]{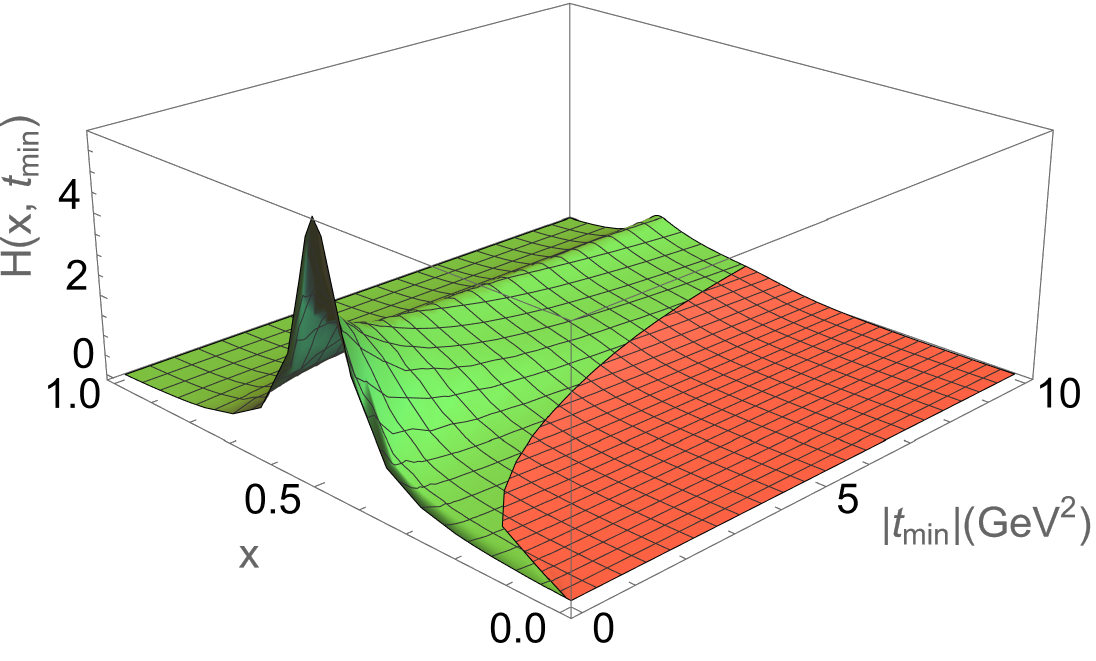}
   \caption{GPD $H(x,\zeta(t))$ (top) and $H(x,t(\zeta))$ (bottom) for ${\bm\Delta}_\perp=0$, 
where $\zeta$ and $t$ are related by $-t = \zeta^2 M_T^2 / (1 - \zeta)$.}
  \label{fig:GPDHezet}
\end{figure}

Figure~\ref{fig:PDF} shows the PDF of a He target, compared to the result in (1+1)D~\cite{CCJO}.
We observe that the PDF in (1+1)D is more sharply peaked at the center $(x = 1/2)$  and more strongly suppressed at the end points than its (3+1)D counterpart.
This sharper peak arises from the analytic structure of the denominator in the (1+1)D expression and the absence of transverse momentum 
degrees of freedom, which enhances endpoint suppression. 
In contrast, in (3+1)D, the integration over transverse momentum \(\bm{k}_\perp\) spreads the probability distribution more broadly in \(x\), 
reducing the peak at \(x = 1/2\) and increasing the relative weight near the end points.
In particular, for the weakly bound He-like state considered here, the PDF shows a nonrelativistic, peak-dominated shape centered at 
$x=1/2$, similar to the peaking-approximation trend most pronounced in (1+1)D. 
As the binding energy increases, this peak broadens, reflecting the stronger relativistic spread in the LF momentum distribution.

Figure~\ref{fig:GPDHe} shows the three-dimensional (3D, top) and  two-dimensional (2D, bottom) GPDs at fixed $-t=0.1$ GeV$^2$ 
obtained from Eq.~\eqref{eq:GPDnonperp}, i.e., in the ${\bm\Delta}_\perp\neq 0$ frame.
The green and red regions in the 3D plot shows the contributions from 
the $H_{\rm DGLAP}(x,\zeta,t)$ and the $H_{\rm ERBL}(x,\zeta,t)$, respectively. The corresponding GPDs in the 2D plot
are shown by solid and dashed lines, respectively.

Our model satisfies the continuity of DGLAP and ERBL GPDs at $x=\zeta$. 
We also note that in the ERBL region ($0 < x < \zeta$) the GPD takes negative values over part of the $x$ range. 
Such sign changes are physically admissible since the ERBL domain describes 
$q\bar{q}$ correlation amplitudes\footnote{Here, ``correlation amplitude'' refers to the LF overlap 
between the initial hadron state and a final state containing the hadron plus a $q\bar{q}$ pair, 
i.e., the emission and subsequent reabsorption of the pair. 
As an amplitude rather than a probability density, it may take negative values due to interference 
between different LF time-ordered contributions.} 
rather than probability densities, and arise naturally from interference between valence and non-valence contributions, 
while all formal constraints such as continuity at $x=\zeta$ and polynomiality are preserved, as discussed later. 

If we take ${\bm\Delta}_\perp=0$, the variables $\zeta$ and $t$ are no longer independent but related through 
$-t = |t_{\rm min}| = \zeta^2 M_T^2 / (1 - \zeta)$. 
In this case, the GPD can be written as $H(x,\zeta(t))$ or equivalently $H(x,t(\zeta))$ over the full range 
$0 \le \zeta(t) \le 1$, corresponding to $0 \le -t < \infty$, 
as given in Eq.~\eqref{eq:ERBLzeta}. 
The results for $H(x,\zeta(t))$ (top) and $H(x,t(\zeta))$ (bottom) are shown in Fig.~\ref{fig:GPDHezet}.

\begin{figure}
  \includegraphics[width=\linewidth]{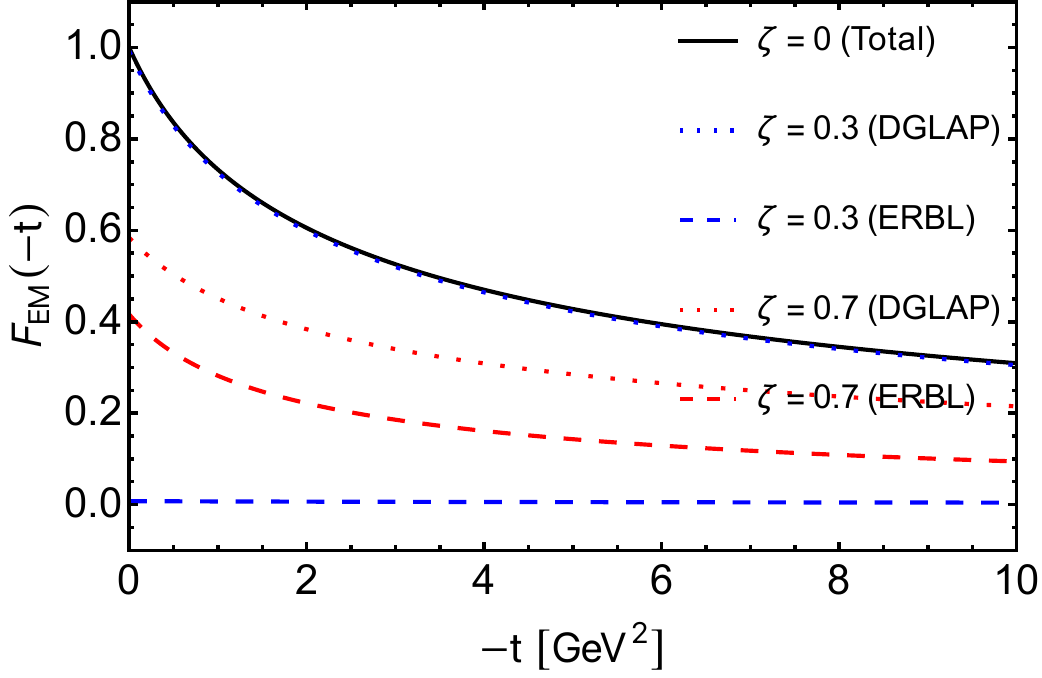}
   \caption{The valence (DGLAP) and non-valence (ERBL) contributions to the first Mellin moment (EM form factor) for $\zeta = 0$, $0.3$, and $0.7$.}
  \label{fig:EMformvalnv}
\end{figure}

In Fig.~\ref{fig:EMformvalnv}, we show the valence (DGLAP, dotted) and non-valence (ERBL, dashed) contributions 
to the first Mellin moment (EM form factor, solid line) for $\zeta = 0$, $0.3$, and $0.7$. 
For $\zeta = 0$, the result arises solely from the valence contribution and coincides with the total. 
For nonzero $\zeta$, the sum of the valence and non-valence contributions reproduces the total result.

As discussed earlier, for analyzing the polynomiality of higher Mellin moments, the symmetric variables $(X,\xi)$ are more suitable than the asymmetric $(x,\zeta)$ variables.
Using these symmetric variables, we replot the 3D GPD $H(X,\xi,t)$ in Fig.~\ref{fig:GPDSym},
where $H_{\rm DGLAP}(X,\xi,t)$ corresponds to the ranges $\xi \leq X \leq 1$ and $-1 \leq X \leq -\xi$,
and $H_{\rm ERBL}(X,\xi,t)$ corresponds to the range $-\xi \leq X \leq \xi$.
This GPD satisfies the polynomiality conditions given by Eq.~\eqref{eq:Poly1}.

\begin{figure}
  \includegraphics[width=\linewidth]{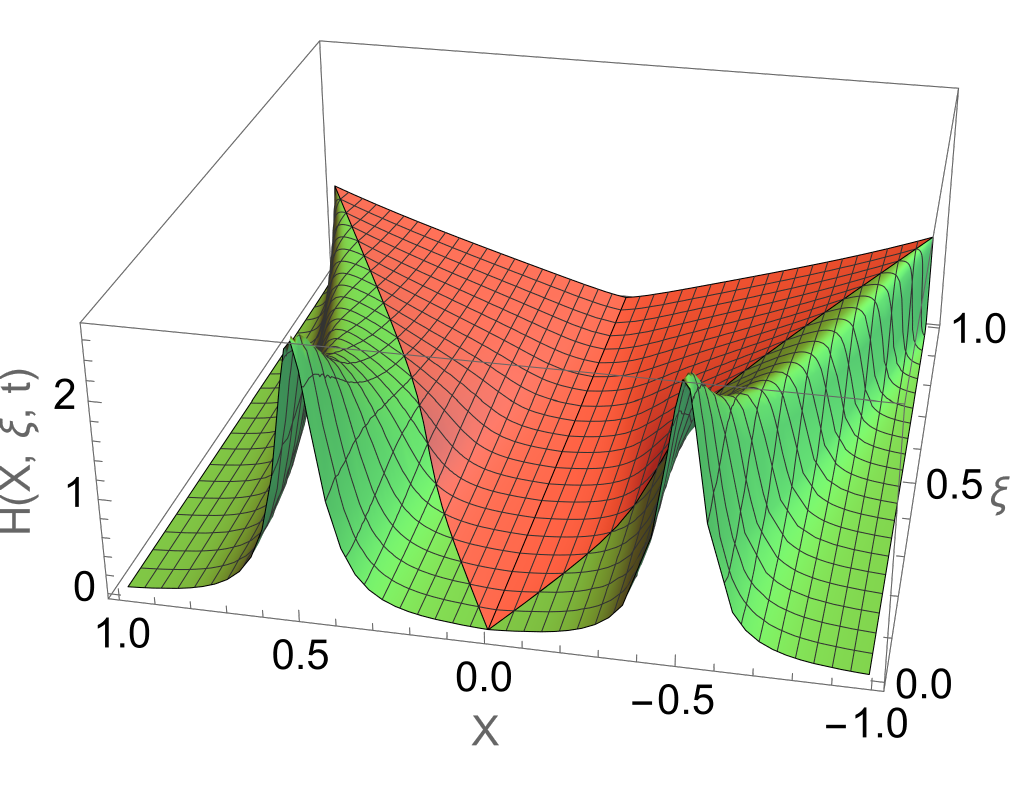}
   \caption{3D GPD $H(X,\xi,t)$ in symmetric $(X, \xi)$ variables.}
  \label{fig:GPDSym}
\end{figure}
\begin{figure}
    \includegraphics[width=\linewidth]{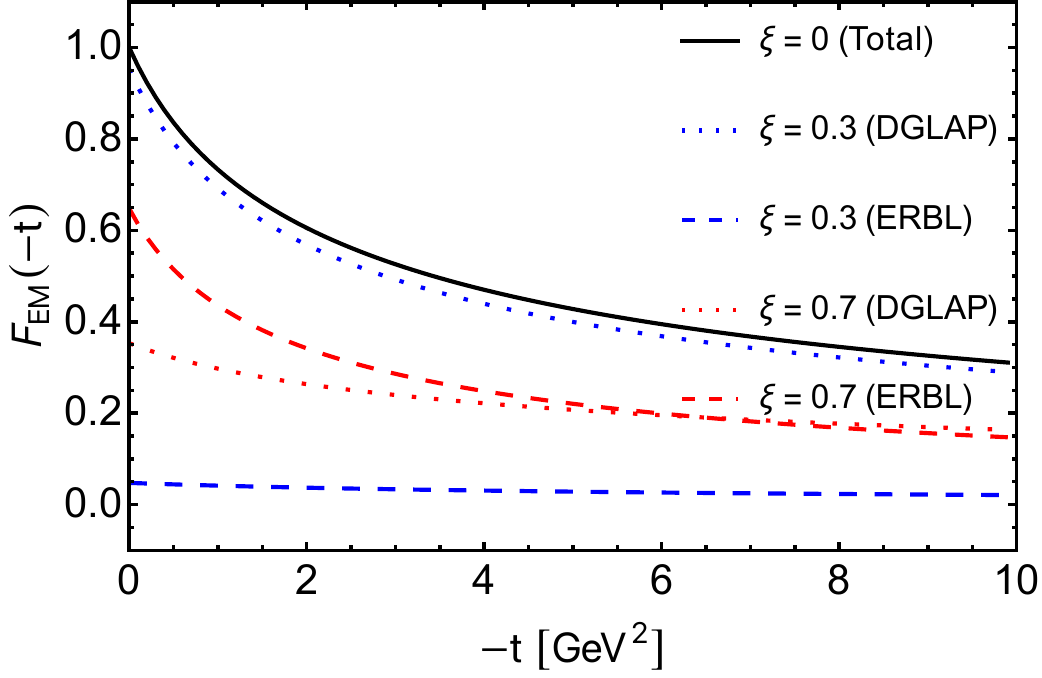}
    \includegraphics[width=\linewidth]{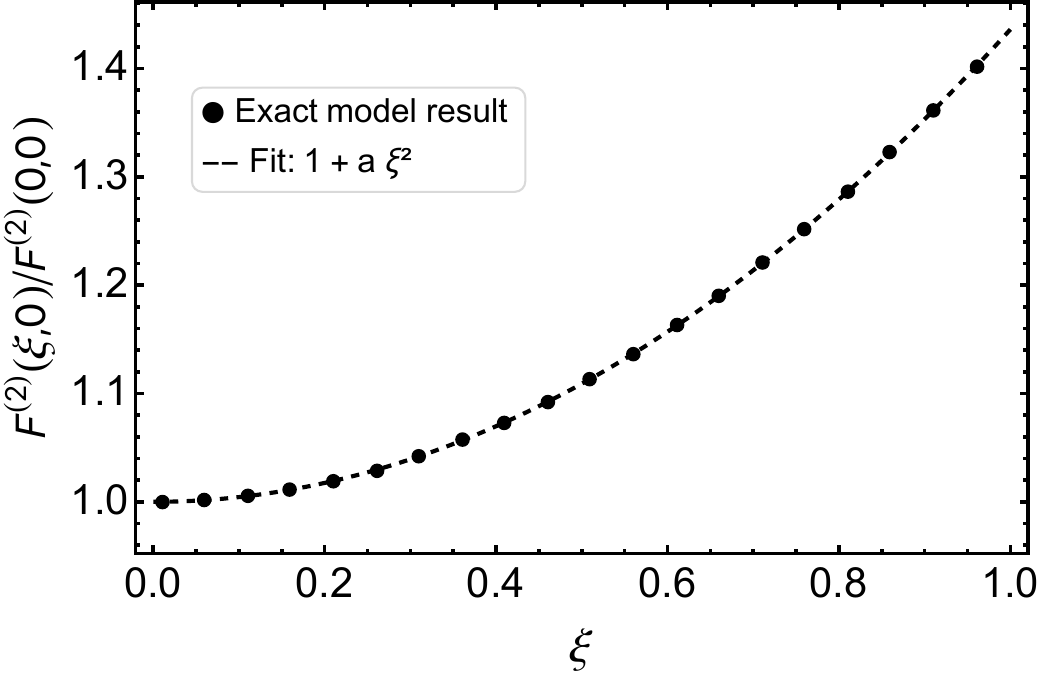}
    \caption{First Mellin moment ($n=1$), corresponding to the EM form factor (top), as a function of $t$, and the normalized second Mellin moment ($n=2$, bottom) at fixed $t=0$ as a function of the symmetric variable $\xi$.}
    \label{fig:Polynomial}
\end{figure}
Figure~\ref{fig:Polynomial} shows the first $(n=1)$ Mellin moment (EM form factor, top) 
and the normalized second $(n=2)$ Mellin moment $F^{(2)}(\xi,t)$ (bottom) of the GPD. 
While the EM form factor obtained from the $(X,\xi)$ variables is essentially 
identical to that from the $(x,\zeta)$ variables, the individual DGLAP and ERBL contributions differ in appearance. 
These vanish once the $\xi$-$\zeta$  relation is taken into account.

For the second moment, we plot the normalized quantity 
$F^{(2)}(\xi, 0) / F^{(2)}(0, 0)$ as a function of $\xi$ at fixed $t=0$, namely
\be\label{eq:F2norm}
\frac{F^{(2)}(\xi,0)}{F^{(2)}(0,0)} = 1 + \xi^2 \frac{A_{22}(0)}{A_{20}(0)} \equiv 1 + a(0)\,\xi^2,
\ee
which provides a direct test of the polynomiality relation in Eq.~\eqref{eq:Poly2}.  
We obtain $a(0)\simeq 0.436$, equal to the ratio $A_{22}(0)/A_{20}(0)$.  
More generally, the normalized second moment can be written in the same polynomial form,  
$\frac{F^{(2)}(\xi,t)}{F^{(2)}(0,t)} = 1 + a(t)\,\xi^2$,
where the slope depends on the value of $t$, implying the relation   $A_{20}(t)=a(t) A_{22}(t)$. 
The resulting generalized form factors $A_{20}(t)$ and $A_{22}(t)$ for the $^4{\rm He}$ target are displayed in Fig.~\ref{fig:GFF}.  

In this model calculation, the values at $t=0$ are found to be  
\be
A_{20}(0)=1, \qquad A_{22}(0)=0.436.
\ee
\begin{figure}
    \includegraphics[width=\linewidth]{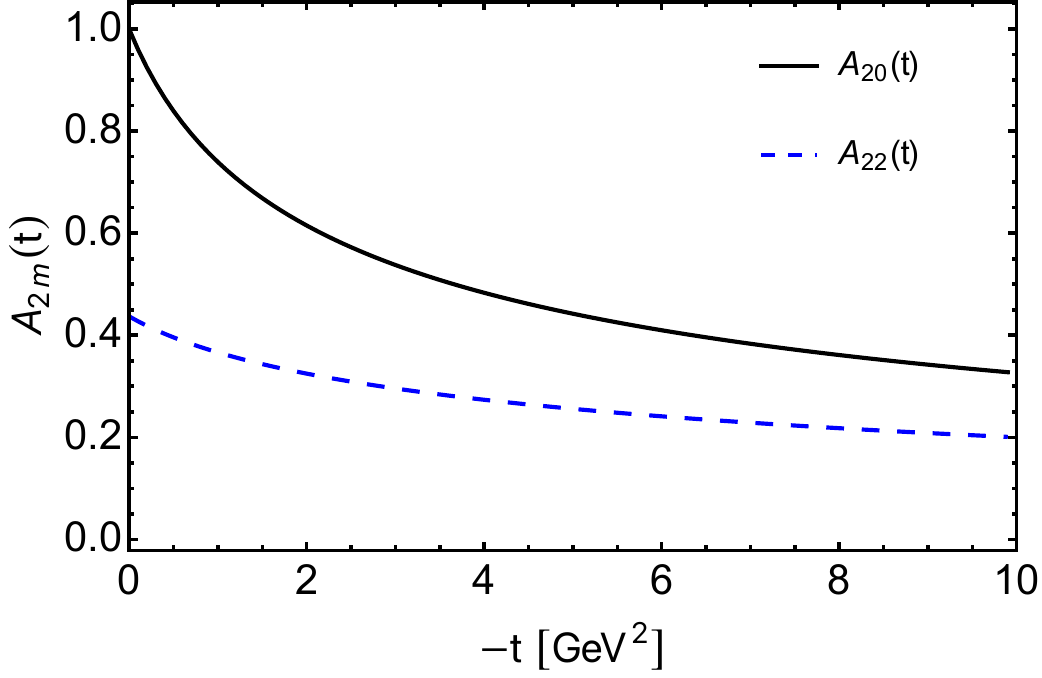}
    \caption{Generalized form factors $A_{20}(t)$ and $A_{22}(t)$.}
    \label{fig:GFF}
\end{figure}

In our spin-0 (purely scalar) model, the mapping of the second Mellin moment to the GFFs may involve the $\xi^{2}$ coefficient 
entering with a minus sign in the $D$-term, in which case one could write
$D(t) = -c A_{22}(t)$ with $c>0$,
so that a positive $A_{22}(t)$ would correspond to a negative $D(t)$. 
The precise value of $c$ and the exact sign relation can depend on the normalization 
conventions for the Mellin moments and the energy-momentum tensor (EMT). 
As a reference, the authors in~\cite{HS17} reports $D(t)=-1$ for a free spin-0 (Klein-Gordon) theory, and less negative values in strongly
interacting cases for $h=\pi,K,\eta$, e.g., $D_{\pi}=-0.91(1)$, $D_{K}=-0.77(15)$, $D_{\eta}=-0.69(19)$.

To fully verify $c$ and the sign of the $D$-term in our framework, however, we will determine the 
GFFs directly from the EMT matrix element calculation in a separate work.

\subsection{CFFs and BSA}
\begin{figure}
  \includegraphics[width=\linewidth]{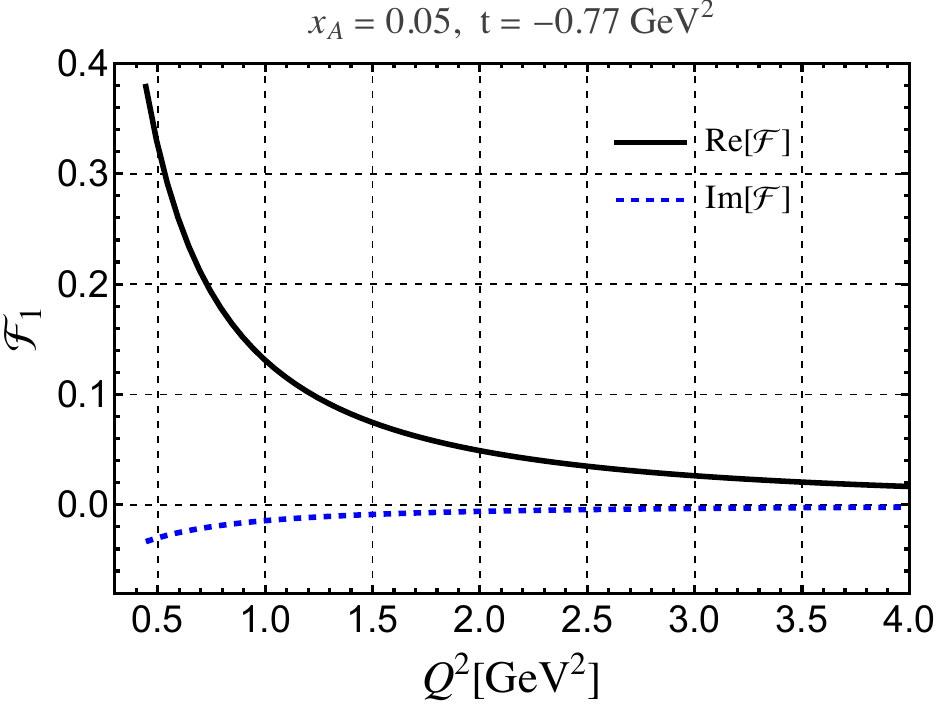}
  \includegraphics[width=\linewidth]{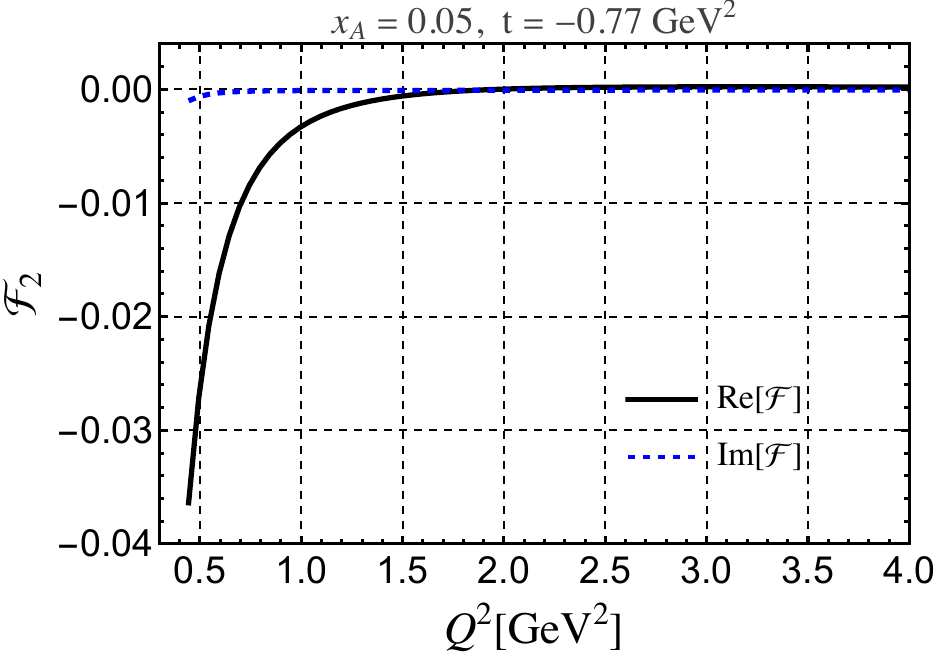}
   \caption{Real and imaginary parts of two Compton form factors for given $x_{A}=0.05$ and $t=-0.77 {\rm GeV^{2}}$.}
  \label{fig:CFF1}
\end{figure}
\begin{figure}
  \includegraphics[width=\linewidth]{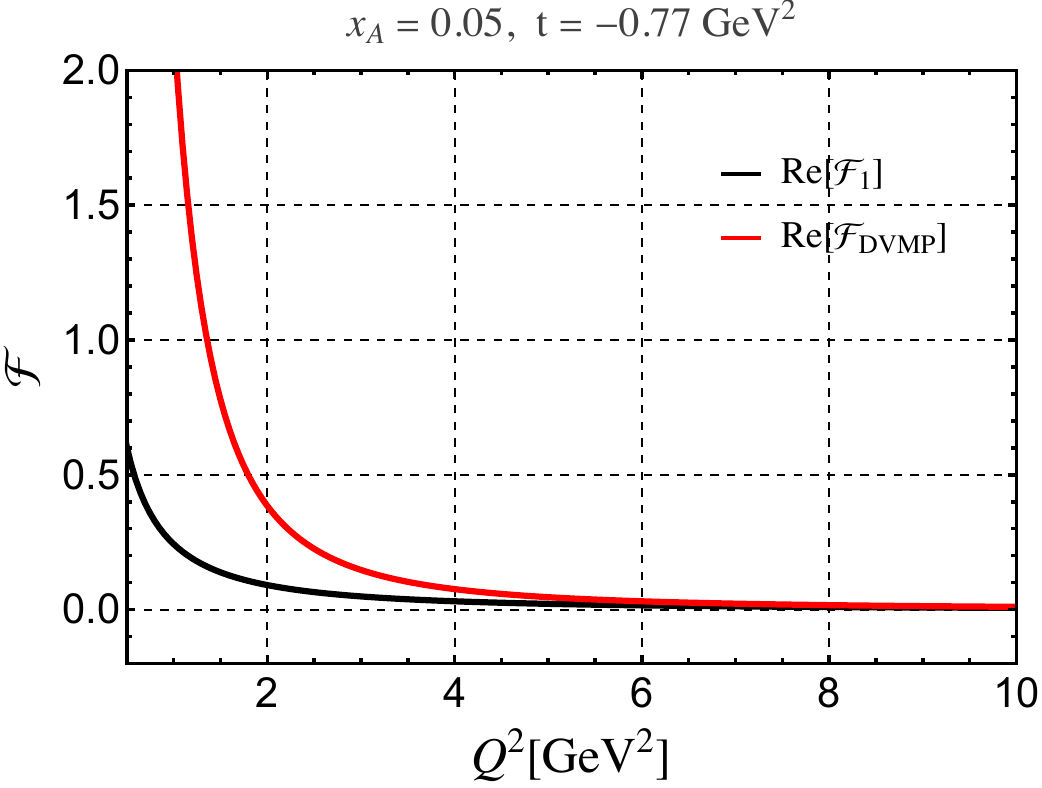}
  \includegraphics[width=\linewidth]{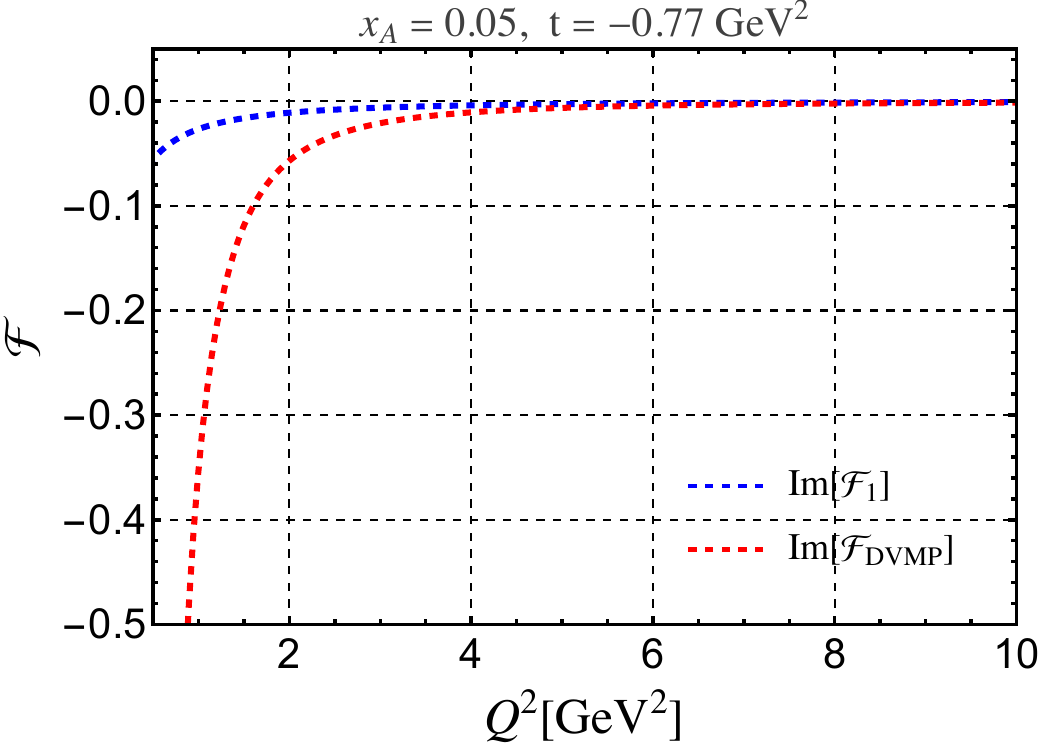}
   \caption{Comparisons of the real and imaginary parts of the Compton form factors obtained from the full calculation and DVMP limit for the given $x_{A}=0.05$ and $t=-0.77 {\rm GeV^{2}}$.}
  \label{fig:CFF2}
\end{figure}

Using the same normalization constant ${\cal N}$ fixed from the PDF normalization, we compute the two CFFs. 

Figure~\ref{fig:CFF1} shows the real (solid line) and imaginary (dashed line) parts of the exact CFFs  
${\cal F}_1$ (top) and ${\cal F}_2$ (bottom) from the VMP process as functions of $Q^2$ for $x_{A} = 0.05$ and $t = -0.77~\mathrm{GeV}^2$.
For these kinematics, the smallest accessible value of $Q^{2}$ is $Q^2_{\rm min}\approx 0.445~\mathrm{GeV}^2$.
As seen in the figure, the magnitude of ${\cal F}_1$ is larger than that of ${\cal F}_2$ 
by roughly an order of magnitude at small $Q^{2}$, 
and their relative difference decreases with increasing $Q^2$. 
Since the BSA is nonzero when at least the imaginary part of either ${\cal F}_1$ or ${\cal F}_2$ is nonzero, 
a finite BSA is obtained up to intermediate $Q^2$. 
In this kinematic regime, smaller $Q^2$ values are more favorable for producing a sizable BSA than larger $Q^2$ values.
In the DVMP limit, only a single CFF, ${\cal F}_{\rm DVMP}$, corresponding to the CFF ${\cal F}_1$, can be obtained. 
Figure~\ref{fig:CFF2} compares the real (top) and imaginary (bottom) parts of the exact CFF ${\cal F}_1$ 
with those of ${\cal F}_{\rm DVMP}$. As $Q^{2}$ increases, the difference between ${\cal F}_1$ and ${\cal F}_{\rm DVMP}$ become progressively smaller.

\begin{figure}
  \includegraphics[width=\linewidth]{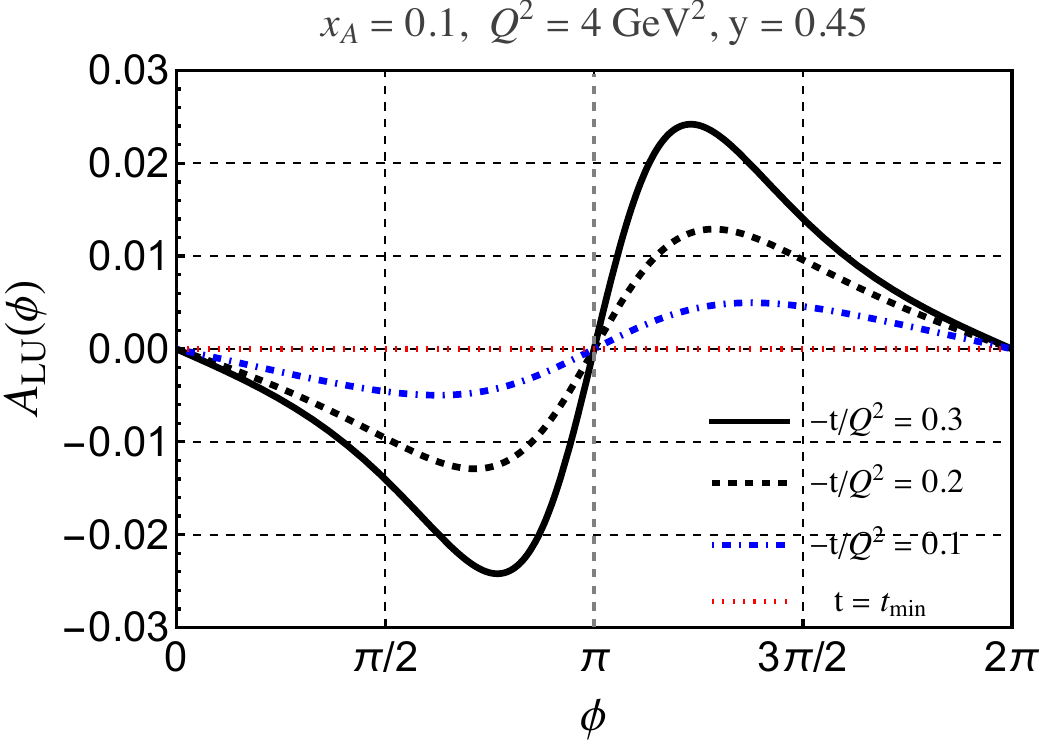}
   \caption{Azimuthal angle $\phi$ dependence of the beam spin asymmetry for various $t$ values (in ${\rm GeV}^{2}$) at fixed $x_{A}=0.1$, $Q^{2}=4 {\rm GeV}^{2}$, and $y=0.45$.}
  \label{fig:BSA}
\end{figure}

Once the two CFFs are obtained for a given set of kinematical parameters, they can be directly related to the BSA, 
as outlined in Ref.~\cite{CCLB}. Fig.~\ref{fig:BSA} presents the BSA, 
${\rm A_{LU}}(\phi)=\frac{d\sigma_{\lambda=+1} - d\sigma_{\lambda=-1}}{d\sigma_{\lambda=+1} + d\sigma_{\lambda=-1}}$ 
as defined in Eq.\eqref{BSA_scalar}, for various values of $t$ at $x_{A}=0.1$, $Q^{2}=4~{\rm GeV}^{2}$, 
and $y=0.45$. Here, the inelasticity $y=\frac{p\cdot q}{p\cdot k}$ denotes the fraction of the incoming electron's energy 
transferred to the target in the target rest frame. The solid, dashed, and dot-dashed
corresponds to $-t/Q^{2}= 0.3,~0.2$, and 0.1, respectively.
As $-t/Q^{2}$ decreases, the magnitude of BSA decreases, vanishing entirely when $t = t_{\rm min}$ (dotted line).

\section{CONCLUSION and Outlook}
In this work, we investigated the beam spin asymmetry in the context of exclusive scalar meson production off a scalar target nucleus,
extending the previous (1+1)-dimensional model analysis~\cite{CCJO} to the present (3+1)-dimensional model analysis. 
This exploration of computing the BSA illuminated some of the ``bare-bone" structure of quantum chromodynamics involved in the scattering process, providing valuable background information for QCD phenomenology. 

Our primary finding is that the BSA is non-vanishing for scalar meson production. This result is significant because it contrasts with the prediction of the leading-twist GPD formulation, which suggests a vanishing BSA due to the reduction of the correlation amplitudes in the asymptotic region where the factorization of the GPD is attainable. Without the BH-Compton interference in the squared amplitude, the underlying hadronic correlation amplitudes could be accessed with minimal contamination. The discrepancy in predicting the BSA highlights the complementary nature of our work to the GPD formulation, indicating that the GPD formulation may not fully capture the complexities of the process. 

By exploring different $-t/Q^2$ regions, we could summarize the applicability of the leading-twist GPDs. Our results show that the leading-twist GPD formulation is valid for the Helium-4 target in 3+1D when $-t/Q^2$ is on the order of $10^{-2}$ and smaller. This finding has implications for the ongoing research at facilities like the Jefferson Lab 12 GeV electron beam and the future Electron-Ion Collider (EIC), suggesting that the EIC may be better suited to access the forward angle kinematics necessary for the GPD formulation to be valid. 

Building on this work, future research could explore the next-to-leading order GPD first moment, which is related to the gravitational form factor. By extending our model to include a derivative-type coupling, we could calculate the energy-momentum tensor and access the mass form factor A, spin form factor B, and the D form factor related to the deformation of the target. It would be particularly interesting to determine the sign of the D term under this extended model calculation, as it has implications for the stability of the proton. 

Another potential avenue for future investigation is the inclusion of fermion loops in our calculations. This extension would allow us to explore not only the chiral-even leading-twist GPD but also the chiral-odd GPD. It would be also interesting to access the chiral-odd GPD  investigating a derivative coupling at the meson production vertex. If the GPD prediction then provides a non-vanishing BSA, it would suggest a significant contribution from the chiral-odd GPD that overcomes the suppressing $\sqrt{-t}/Q$ factor. Further extensions of our work could involve TMD computations and explorations of the light-front wavefunctions from the analyses of the light-front quark model. 

\section*{Acknowledgements}
This work was supported in part by the U.S. Department of Energy (Grant No. DE-FG02-03ER41260). 
The National Energy Research Scientific Computing Center (NERSC) supported by the Office of Science of the U.S. Department of Energy 
under Contract No. DE-AC02-05CH11231 is also acknowledged. 
Y. Choi and H.-M. Choi were supported by the National Research Foundation of Korea (NRF) grant funded by the Korea government (MSIT) 
under Grant No. RS-2025-02634319 (Y. Choi) and RS-2023-NR076506 (H.-M. Choi). 

\bibliographystyle{apsrev4-2}
\bibliography{references}
\end{document}